\documentclass[runningheads]{llncs} 
\pdfoutput=1
\usepackage[utf8]{inputenc}
\usepackage{algorithm} 
\usepackage{algpseudocode}  
\usepackage[english]{babel}
\usepackage{color}
\usepackage{colortbl,tabulary,tabularx,multirow}
\usepackage{ltxtable}
\usepackage{comment,paralist}
\usepackage{wrapfig}
\usepackage{amsmath}
\usepackage{amssymb}
\usepackage{stmaryrd}
\usepackage{braket}
\usepackage{textcomp}
\usepackage{listings}
\usepackage{theorem}
\usepackage{alltt}
\usepackage{epsfig,psfrag,subfigure}
\usepackage{graphicx}
\usepackage[colorlinks,linkcolor=blue,citecolor=blue,urlcolor=blue]{hyperref}
\usepackage{multicols}
\usepackage{balance}
\usepackage{amssymb}
\newcommand{\changefont}[4]{\fontencoding{#1}\fontfamily{#2}\fontseries{#3}\fontshape{#4}\selectfont}

\newcommand{\R}{\mathbb{R}}
\newcommand{\N}{\mathbb{N}}

\everymath{\displaystyle}

\usepackage{tikz}
\usetikzlibrary{shapes,arrows,decorations.pathmorphing,backgrounds,fit,matrix,calc}

\newcommand{\krons}{\otimes_{sym}}

\definecolor{todo-color}{rgb}{1,0,0}

\definecolor{ploc-color}{rgb}{0.9,0.6,0}

\definecolor{romain-color}{rgb}{1,0,1}

\definecolor{pierre-color}{rgb}{0,1,1}

\definecolor{no-color}{rgb}{1,0.2,0.2}

\definecolor{MyDarkGreen}{rgb}{0.0,0.4,0.0}
\definecolor{MyBlue}{rgb}{0.0,0.0,0.7}
\definecolor{MyPurple}{rgb}{0.7,0.0,0.7}
\newcommand{\annotf}[1]{{\changefont{T1}{cmr}{b}{n}\color{MyDarkGreen}\small#1}} 
\newcommand{\annotfc}[1]{{\changefont{T1}{cmr}{b}{n}\color{MyBlue}\bfseries#1}} 
\newcommand{\annotc}[1]{{\changefont{T1}{cmr}{bx}{n}\small \bfseries#1}} 
\newcommand{\annotcf}[1]{{\changefont{T1}{cmr}{b}{n}\color{MyPurple}\small#1}} 

\newcommand{\lsqr}[1]{\operatorname{lsqr}\left(#1 \right)}

\newcommand{\benum}{\begin{enumerate}}
\newcommand{\eenum}{\end{enumerate}}
\newcommand{\nc}{\newcommand}
\newcommand{\rnc}{\renewcommand}
\newcommand{\bvb}{\begin{verbatim}}
\newcommand{\evb}{\end{verbatim}}

\newcommand{\Tr}[1]{\operatorname{Tr}\left(#1 \right)}
\renewcommand{\det}[1]{\operatorname{det}\left(#1\right)}
\DeclareMathOperator{\Dg}{Diag}
\DeclareMathOperator{\size}{length}
\DeclareMathOperator{\return}{return}
\DeclareMathOperator{\vecs}{vecs}
\DeclareMathOperator{\mats}{mats}

\nc{\<}{\langle}
\rnc{\>}{\rangle}
\nc{\bq}{\textbf}
\nc{\m}{\textrm}
\nc{\bb}{\mathbb}
\nc{\til}{\texttildelow}
\nc{\be}{\begin{equation}}
\nc{\ee}{\end{equation}}
\nc{\dps}{\displaystyle}
\rnc{\l}{\left(}\rnc{\r}{\right)}
\nc{\lc}{\left\{}\nc{\rc}{\right\}}
\nc{\lb}{\left[}\nc{\rb}{\right]}
\nc{\ba}[1]{\begin{array}{#1}}
\nc{\ea}{\end{array}}       
\nc{\ra}{\rightarrow}
\nc{\li}{\left |}
\nc{\ri}{\right |}
\nc{\pde}[2]{\frac{\partial #1}{\partial #2}}
\nc{\ode}[2]{\frac{d #1}{d #2}}
\nc{\odee}[3]{\frac{d^{#3} #1}{d #2^{#3}}}
\nc{\pdee}[3]{\frac{\partial^{#3} #1}{\partial #2^{#3}}}
\nc{\bn}{\begin{enumerate}}
\nc{\en}{\end{enumerate}}
\nc{\bt}{\begin{theorem}}
\nc{\et}{\end{theorem}}
\nc{\y}[1]{\lambda_{#1}}
\nc{\ninf}{{\oplus}^{-\infty}}
\nc{\pinf}{{\oplus}^{+\infty}}
\nc{\nninf}{{\otimes}^{-\infty}}
\nc{\ppinf}{{\otimes}^{+\infty}}
\nc{\ir}{\mathbb{I}\mathbb{R}}

\nc{\ep}{\mathcal{E}_{P}}
\nc{\mr}{\mathcal{M}_{r}}
\nc{\mfa}{\mathcal{M}_{f,a}}
\nc{\mfp}{\mathcal{M}_{f,p}}
\nc{\mt}{\m{T}}
\nc{\F}{\mathbb{F}}

\tikzstyle{block} = [draw,rectangle,thick,minimum height=2em,minimum width=\textwidth]
\tikzstyle{sum} = [draw,circle,inner sep=0mm,minimum size=2mm]
\tikzstyle{connector} = [->,thick]
\tikzstyle{line} = [thick]
\tikzstyle{branch} = [circle,inner sep=0pt,minimum size=1mm,fill=black,draw=black]
\tikzstyle{guide} = []
\tikzstyle{snakeline} = [connector, decorate, decoration={pre length=0.2cm,
                         post length=0.2cm, snake, amplitude=.4mm,
                         segment length=2mm},thick, magenta, ->]

\usepackage{environ}

\def\comment#1{}
\newlength{\hsbw}

\tikzstyle{mybox} = [draw, very thick, rectangle, rounded corners, inner sep=0pt, inner ysep=2pt]
\tikzstyle{fancytitle} =[fill=white, draw, rectangle, rounded corners, very thick]
\newsavebox{\GrayRoundedBox}
\NewEnviron{mytikzbox}[1]{%
  \begin{lrbox}{\GrayRoundedBox}
    \setlength{\hsbw}{\linewidth}
    \addtolength{\hsbw}{-8pt}
    \begin{minipage}[b]{\hsbw}
      \begingroup\small
      \begin{alltt}
        \BODY
      \end{alltt}
      \endgroup
    \end{minipage}
  \end{lrbox}
  \begin{flushleft}
    \vspace{-10pt}
    \begin{tikzpicture}
      \node[mybox](box){\usebox{\GrayRoundedBox}};
      \node[fancytitle, left=10pt] at (box.north east) {\tiny {#1}};
    \end{tikzpicture}
  \end{flushleft}
}

\newsavebox{\Zname}
\tikzset{diagram background/.style={fill=blue!5,rounded corners=0.5cm}}
\tikzstyle{block} = [rectangle, draw, fill=blue!20, 
    minimum width=4em, 
    text centered, rounded corners, minimum height=3em]
\tikzstyle{proof block} = [rectangle, draw, fill=magenta!10,
    minimum width=4.2em, 
     rounded corners, minimum height=1.5em]
\tikzstyle{frama block} = [rectangle, draw, fill=green!20, minimum width = 6em,
  minimum height = 4em, rounded corners]
\tikzstyle{library block} = [rectangle, draw, fill=gray!20, minimum width = 4em,
  minimum height = 2em, rounded corners]
\tikzstyle{invisible} = [opacity=0] 
\tikzstyle{every picture}+=[remember picture]

\begin{document}
\mainmatter

\title{Credible Autocoding of Convex Optimization Algorithms} 

\titlerunning{}

\author{Timothy Wang\inst{1} \and Romain Jobredeaux\inst{1} \and Marc Pantel\inst{4} \and Pierre-Loic Garoche\inst{2} \and  Eric Feron\inst{1} \and Didier Henrion \inst{3} }

\authorrunning{ }

\institute{Georgia Institute of Technology,
  Atlanta, Georgia, USA \and ONERA -- The French Aerospace Lab, Toulouse, FRANCE
\and CNRS-LAAS -- Laboratory for Analysis and Architecture of Systems, Toulouse, FRANCE 
\and ENSEEIHT, Toulouse, FRANCE}

\maketitle

\begin{abstract}
The efficiency of modern optimization methods, coupled with increasing computational resources, 
has led to the possibility of real-time optimization algorithms acting in safety critical roles. 
There is a considerable body of mathematical proofs on on-line optimization programs which
can be leveraged to assist in the development and verification of their implementation.
In this paper, we demonstrate how theoretical proofs of real-time optimization algorithms can
be used to describe functional properties at the level of the code, thereby making 
it accessible for the formal methods community. 
The running example used in this paper is a generic semi-definite programming (SDP) solver.
Semi-definite programs can encode a wide variety of optimization problems and can be solved in polynomial time at a given accuracy.
We describe a top-to-down approach that transforms a high-level analysis of the algorithm into useful code annotations.
We formulate some general remarks about how such a task can be incorporated into a convex programming autocoder. 
We then take a first step towards the automatic verification of the optimization program by identifying key issues to be adressed 
in future work.
\end{abstract}

\section{Introduction}

The applications of optimization algorithms are not only limited to large scale, off-line problems on the desktop. 
They also can perform in a real-time setting as part of safety-critical systems in control, guidance and navigation. 
For example, modern aircrafts often have redundant control surface actuation, which allows the
possibility of reconfiguration and recovery in the case of emergency. The precise re-allocation of the actuation resources 
can be posed, in the simplest case, as a linear optimization problem that needs to be solved in real-time. 

In contrast to off-line desktop optimization applications, real-time embedded optimization code needs
to satisfy a higher standard of quality, if it is to be used within a safety-critical system. 
Some important criteria in judging the quality of an embedded code include the predictability of 
its behaviors and whether or not its worst case computational time can be bounded.
Several authors including Richter~\cite{richter13}, Feron and McGovern~\cite{lmcgovern}\cite{mcgovern_feron}
have worked on the certification problem for on-line optimization algorithms used in control, in
particular on worst-case execution time issues. 
In those cases, the authors have chosen to tackle the problem at a high-level of abstraction. 
For example, McGovern reexamined the 
proofs of computational bounds on interior point methods for semi-definite programming; 
however he stopped short 
of using the proofs to analyze the implementations of interior point methods. 
In this paper, we extend McGovern's work further by demonstrating the expression of 
the proofs at the code level for the certification of on-line optimization code.
The utility of such demonstration is twofolds. First, we are considering the reality that 
the verifications of safety-critical systems are almost always done at the source code level. 
Second, this effort provides an example output that is much closer to being an 
accessible form for the formal methods community. 

The most recent regulatory documents such as DO-178C~\cite{do178c} and, in particular, 
its addendum DO-333~\cite{do333}, advocate the use of formal methods in the verification 
and validation of safety critical software.
However, complex computational cores in domain specific software such as control or optimization
software make their automatic analysis difficult in the absence of input from domain experts. It
is the authors' belief that communication between the communities of formal software analysis and 
domain-specific communities, such as the optimization community, are key to successfully express
the semantics of these complex algorithms in a language compatible with the application of
formal methods. 

The main contribution of this paper is to present the expression, formalization, and translation of
high-level functional properties of a convex optimization algorithm along with their proofs down 
to the code level for the purpose of formal program verification. 
Due to the complexity of the proofs, we cannot yet as of this moment, 
reason about them soundly on the implementation itself. 
Instead we choose an intermediate level of abstraction of the implementation where
floating-point operations are replaced by real number algebra. 

The algorithm chosen for this paper is based on a class of optimization 
methods known collectively as interior point methods.
The theoretical foundation behind modern interior point methods
can be found in Nemrovskii et. al~\cite{Nemirovskii88}\cite{Nemirovskii89}. 
The key result is the self-concordance of certain barrier functions that guarantees the 
convergence of a Newton iteration to an $\epsilon$-optimal solution in polynomial time. 
For more details on polynomial-time interior point methods, readers can refer to~\cite{nesterov94}. 

Interior-point algorithms vary in the 
Newton search direction used, the step length, the initialization process, and whether or 
not the algorithm can return infeasible answers in the intermediate iterations. 
Some example search directions are the 
Alizadeh-Haeberly-Overton (AHO) direction~\cite{Alizadeh94primal-dualinterior-point}, the Monteiro-Zhang (MZ)
directions~\cite{Monteiro97}, the Nesterov-Todd (NT) direction~\cite{Nesterov95primal}, and 
the Helmberg-Kojima-Monteiro (HKM) direction~\cite{Helmberg96aninterior-point}. 
It was later determined in~\cite{Monteiro98aunified} that all of these search directions can be captured by a particular scaling matrix in 
the linear transformation introduced by Sturm and Zhang in~\cite{zhang98}.  
An accessible  introduction to 
semi-definite programming using interior-point method can be found in the works of Boyd 
and Vandenberghe~\cite{boydcvx04}. 

Autocoding is the computerized process of translating the specifications of an algorithm, 
that is initially expressed in a high-level modeling language such as Simulink, 
into source code that can be transformed further into an embedded executable binary. 
An example of an autocoder for optimization programs can be found in the work of Boyd~\cite{boydcvx}. 
One of the main ideas behind this paper, is that by combining
the efficiency of the autocoding process
with the rigorous proofs obtained from a formal analysis of the optimization
algorithm, we can
create a credible autocoding process~\cite{wjfarxiv2013} 
that can rapidly generate formally verifiable optimization code. 

In this paper, the running example is an interior point algorithm
with the Monteiro-Zhang (MZ) Newton search direction. 
The step length is  fixed to be $1$ and the input problem 
is a generic SDP problem obtained from system and control. 
The paper is organized as follows: first we introduce the basics of program verification and semi-definite programming. 
We then introduce the example interior point algorithm and discuss its known properties. 
We then give a manual
example of a code implementation annotated with the semantics of the optimization algorithm
using the Floyd-Hoare method~\cite{hoareaxiom69}. 
Afterwards, we discuss an approach for automating
the generation of the optimization semantics
and the verification of the generated semantics with respect to the code. 
Finally, we discuss some future directions of research. 

\section{Credible Autocoding: General Principles} 

\begin{figure}[!ht]
\centering
\includegraphics[width=1.0\textwidth]{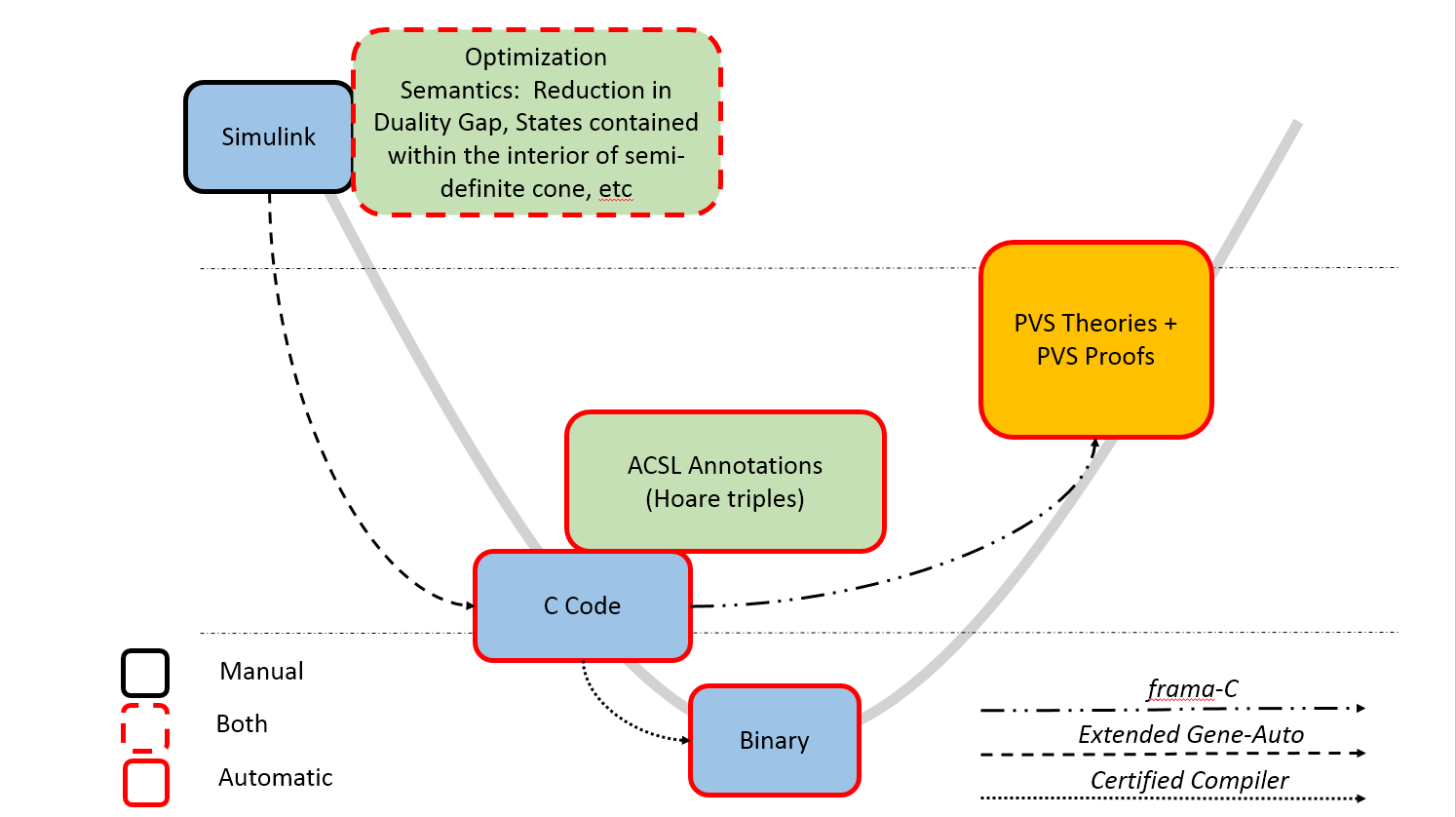}
\caption{Visualization of autocoding and verification process For Optimization Algorithms} 
\label{fig:verified_process}
\end{figure}

In this paper, we introduce a credible autocoding framework for convex optimization algorithms. 
Credible autocoding, analogous to credible compilation from~\cite{rinard99}, is a process by which the 
autocoding process generates formally verifiable evidence that the output source code correctly 
implements the input model. 
An overall view of a credible autocoding framework is given in figure \ref{fig:verified_process}.
Existing work already provides for the automatic generation of embedded convex optimization code~\cite{boydcvx}.
Given that proofs of high-level functional properties of interior point algorithms do exist, we want
to generate the same proof that is sound for the implementation, and expressed in a formal specification language
embedded in the code as comments. 
One of the key ingredients that made credible autocoding applicable for
control systems~\cite{wjfarxiv11} is that the ellipsoid sets generated by synthesizing 
quadratic Lyapunov functions are relatively easy to reason about even on the code level.
The semantics of interior point algorithms, however, do not rely on simple quadratic invariants.
The invariant obtained from the proof of good behavior of interior point algorithms 
is generated by a logarithmic function. 
This same logarithmic function can also be used in showing the optimization algorithm 
terminates in within a specified time. 
This function, is not provided, is perhaps impossible to synthesize from using existing 
code analysis techniques on the optimization source code. In fact nearly all existing
code analyzers only handle linear properties with the notable exception in~\cite{prhscc12}.

\section{Program Verification} 

In this section, we introduce some concepts from program verification that we use later in 
the paper. The readers who are already familiar with Hoare logic and axiomatic semantics should 
skip ahead to the next section. 

\subsection{Axiomatic Semantics} 

One of the classic paradigms in formal verification of programs is the usage of axiomatic semantics. 
In axiomatic semantics, 
the semantics or mathematical meanings of a program is based on the relations between the logic 
predicates that hold true before and after a piece of code is executed. 
The program is said to be partially correct if the logic predicates holds throughout the execution of the program. 
For example, given the simple {\tt while} loop program in figure \ref{loop00}, 
\begin{figure}
\begin{lstlisting}
while (x*x>0.5) 
	x=0.9*x;
end
\end{lstlisting}
\caption{A {\tt while} loop Program}
\label{loop00}
\end{figure}
if we assume the value of variable $x$ belongs to the set $\lb -1, 1\rb$ before the execution of the {\tt while} loop, then
the logic predicate \annotc{x*x<=1} holds before, during and after the execution of the {\tt while} loop.
The predicate that holds before the execution of a block of code is referred to as the pre-condition. 
The predicate that holds after the execution of a block of code is referred to as the post-condition. 
Whether a predicate is a pre or post-condition is contextual since its dependent on the block of code
that its mentioned in conduction with. 
A pre-condition for one line of code can be the post-condition for the previous line
of the code. 
A predicate that remains constant i.e. holds throughout the execution of the program is called
an \emph{invariant}. 
For example, the predicate \annotc{x*x<=1} is an invariant for the {\tt while} loop. 
However the predicate \annotc{x*x>=0.9} is not an invariant since
it only holds during a subset of the total execution steps of the loop. 

The invariants can be inserted into the code as comments. We refer to these comments as code specifications or annotations. 
For example, inserting the predicate \annotc{x*x<=1} into the program in figure \ref{loop00} results in 
the annotated program in figure \ref{loop10}. 
The pseudo Matlab specification language used to express the annotations in figure \ref{loop10} is 
modelled after ANSI/ISO C Specification Language (ACSL~\cite{baudin:acsl}), which is an existing formal specification language for C programs. 
The pre and post-conditions are denoted respectively using ACSL keywords \emph{requires} and \emph{ensures}. 
The annotations are captured within comments denoted by the Matlab comment symbol \%\%. 
\begin{figure}
\begin{lstlisting}
%% requires x*x<=1; 
%% ensures x*x<=1; 
while (x*x>0.5)
	x=0.9*x;
end
\end{lstlisting}
\caption{Axiomatic Semantics for a {\tt while} loop Program}
\label{loop10}
\end{figure}
Throughout the rest of the paper, we use this pseudo Matlab specification language in the annotations of the example 
convex optimization program. 
Other logic keywords from ACSL, such as \emph{exists}, \emph{forall} and \emph{assumes}
are also transferred over and they have their usual meanings. 

\subsection{Hoare Logic} 
We now introduce a formal system of reasoning about the correctness of programs, 
that follows the axiomatic semantics paradigm, called Hoare Logic~\cite{hoareaxiom69}. 
The main structure within Hoare logic is the Hoare triple. 
Let $P$ be a pre-condition for the block of code $C$ and let $Q$ be the post-condition for $C$. We can express 
the annotated program in \ref{loop10} as a Hoare triple denoted by $\lc P \rc C \lc Q \rc$, 
in which both $P$ and $Q$ represent the invariant \annotc{x*x<=1} and $C$ is the {\tt while} loop.
The Hoare triples is \emph{partially correct} 
if $P$ hold true for some initial state $\sigma$, and $Q$ holds 
for the new state $\sigma'$ after the execution of $C$. 
For total correctness, we also need prove termination of the execution of $C$. 

Hoare logic includes a set of axioms and inference rules for reasoning about the correctness of 
Hoare triples for various program structures of a generic imperative programming language. 
Example program structures include loops, branches, jumps, etc.
In this paper, we only consider {\tt while} loops. 
For example, a Hoare logic axiom for the {\tt while} loop is
\be
\frac{ \lc P \wedge B \rc C; \lc P \rc }{ \lc P \rc \m{while } B\m{ do }C\m{ done }\lc \neg B \wedge P \rc}. 
\label{while}
\ee
Informally speaking, the axioms and inference rules should be interpreted as follows: 
the formula above the horizontal line implies the formula below that line. 
From (\ref{while}),  note that the predicate $P$ holds before and after the {\tt while} loop. 
We typically refer to this type of predicate as an \emph{inductive invariant}. 
Inductive invariants require proofs as they are properties
that the producer of the code is claiming to be true. For the {\tt while} loop, according the axiom in (\ref{while}), 
we need to show that the predicate $P$ holds in every iteration of the loop. 
In contrast, axiomatic semantics also allows predicates that are essentially assumptions about the state of the program. 
This is especially useful in specifying properties about the inputs. 
For example, the variable \annotc{x} in figure \ref{loop10} is assumed to have an value between $-1$ and $1$.
The validity of such property cannot be proven since it is an assumption. 
This type of invariant is referred to as an \emph{assertion}. 
In our example, the assertion \annotc{x<=1 \&\& x>=-1} is necessary for proving 
that \annotc{x*x<=1} is an inductive invariant of the loop. 

For this paper, we use some basic 
inferences rules from Hoare logic, which are listed in table~\ref{hoare_rules}. 
\begin{table}
\begin{tabular}{cc} 
	\parbox{5cm}{\be
\frac{\lc P_{1} \Rightarrow P_{2} \rc C \lc Q_{1} \Rightarrow Q_{2} \rc }{\lc P_{1} \rc C \lc Q_{2} \rc}
\label{conseq}
\ee}  &
\parbox{5cm}{\be
	\frac{\lc P \rc C_{1} \lc Q \rc; \lc Q \rc C_{2} \lc W \rc }{\lc P \rc C_1; C_2 \lc W \rc}
	\label{compose}
\ee} \\ 
\parbox{5cm}{\be
\frac{ }{\lc P \rc SKIP \lc P \rc}
\label{skip}
\ee}  &
\parbox{5cm}{\be
	\frac{}{\lc P[e/x] \rc x:=expr \lc P \rc}
	\label{subst}
\ee} \\
\parbox{6cm}{\be
	\frac{}{\lc P \rc x:=expr \lc \exists x_0 \l x=expr\lb x_0 /x \rb\r \wedge P\lb x_0 /x\rb  \ \rc}
	\label{substforward}
\ee}  & \\
\end{tabular}
\caption{Axiomatic Semantics Inference Rules for a Imperative Language}
\label{hoare_rules}
\end{table}
The consequence rule in (\ref{conseq}) is useful whenever a stronger 
pre-condition or weaker post-condition is needed. By stronger, we meant the set defined 
by the predicate is smaller. By weaker, we mean precisely the opposite. 
The substitution rules in (\ref{subst}) and (\ref{substforward}) are used when the code is an assignment statement. 
The weakest pre-condition $P[x/expr]$ in (\ref{subst}) means $P$ with all instances of the expression $expr$ replaced by $x$. 
For example, given a line of code \annotc{y=x+1} and a known weakest pre-condition \annotc{x+1<=1}, 
we can deduct that \annotc{y<=1} is a correct post-condition using the backward substitution rule. 
Although usually (\ref{subst}) is used to compute the weakest pre-condition from the known post-condition. 
Alternatively the forward propagation rule in (\ref{substforward}) is used to compute the 
strongest post-condition. 
The skip rule in (\ref{skip}) is used when executing the piece of code 
does not change any variables in the pre-condition $P$. 

\subsection{Proof Checking} 
The utility of having the invariants in the code 
is that finding the invariants is in general more difficult than checking that given 
invariants are correct. 
By expressing and translating the high-level functional properties and their proofs
onto the code level in the form of invariants, we can verify the correctness of the optimization
program with respect to its high-level functional properties using a proof-checking procedure i.e.
by verifying each use of a Hoare logic rule.

\section{Semi-Definite Programming and the Interior Point Method} 
In this section, we give an overview of the Semi-Definite Programming (SDP) problem. 
The readers who are already familiar with interior point method and convex optimization can 
skip ahead to the next section. 
The notations used in this section are as follows: let $A=(a_{i,j})_{1\leq i,j\leq n},B\in\R^{n \times n}$ be two matrices and $a,b\in\R^n$
be two column vectors. $\Tr{A} = \sum_{i=1}^n a_{i,i}$ denotes the trace of matrix $A$.
$\<\cdot,\cdot\>$ denotes an inner product, defined in $\R^{n \times n}\times\R^{n \times n}$ as $\<A,B\>:= \Tr{B^{\m{T}}A}$ 
and in $\R^{n}\times \R^{n}$ as
$\<a,b\>:=a^{\m{T}} b$. The Frobenius norm of $A$ is defined as $\|A\|_F=\sqrt{\<A,A\>}$.
The symbol $\mathbb{S}^n$ denotes the space of symmetric matrices of size $n\times n$.
The space of $n\times n$ symmetric positive-definite matrices is denoted as 
$\mathbb{S}^{n+}=\left\{S\in\mathbb{S}^n|\forall x\in\R^n\setminus\{0\}, x^\m{T}Sx>0\right\}$. 
If $A$ and $B$ are symmetric, $A\prec B$ (respectively  $A\succ B$) 
denotes the positive (respectively negative)-definiteness of matrix $B-A$.
The symbol $I$ denotes an identity matrix of appropriate dimension. 
For $X,Z \in \mathbb{S}^{n+}$, some basic properties of matrix derivative are 
$\dps \frac{\partial \Tr{XZ}}{\partial X}=Z^{\m{T}}=Z$ and
$\dps \frac{\partial \det{X}}{\partial X}=\det{X}^{-1} \l X^{-1}\r^{\m{T}} = \det{X}^{-1} X^{-1}$. 

\subsection{SDP Problem} 
\label{sec:problem}
Let $n,m\in\N$, $F_0\in \mathbb{S}^{n+}$, $F_1,F_2,\ldots,F_m\in\mathbb{S}^{n}$, 
and $b = \begin{bmatrix} b_1&b_2&\hdots&b_m\end{bmatrix}^\m{T}\in\R^m$.
Consider a SDP problem of the form in (\ref{dual}). 
The linear objective function 
$\<F_{0},Z\>$ is to be maximized 
over the intersection of positive semi-definite cone $\lc Z \in \mathbb{S}^{n} | Z\succeq 0\rc$ and 
a convex region defined by $m$ affine equality constraints. 
\be
\begin{aligned}
& \underset{Z}{\text{sup}}
	& &  \<F_{0}, Z\>, \\
& \text{subject to}
	& &  \<F_{i}, Z\> + b_{i}=0, & i=1,\ldots, m\\
	& &  & Z \succeq 0.  
\end{aligned}
\label{dual}
\ee
Note that a SDP problem
can be considered as a generalization of a linear programming (LP) problem. 
To see this, let $Z=\Dg{\l z \r}$ where $z$ is the standard LP variable. 

We denote the SDP problem in (\ref{dual}) as the \emph{dual} form.   
Closely related to the dual form, is another SDP problem as shown 
in (\ref{prime}), called the \emph{primal} form. 
In the primal formulation, the  
linear objective function $\<b,p\>$ is minimized 
over all vectors $p=\begin{bmatrix} p_1 & \hdots & p_m\end{bmatrix}^\m{T}\in\R^m$
under the semi-definite constraint $F_{0}+\sum_{i=1}^{m} p_{i} F_{i} \preceq 0$.
Note the introduction in (\ref{prime}) of a
variable $X=-F_0-\sum_{i=1}^{m} p_{i} F_{i}$ such that $X\succeq 0$, 
which is not strictly needed to express the problem, but is used in later developments. 
\begin{equation}
\begin{aligned}
	& \underset{p,X}{\text{inf}}
	& &  \<b,p\>\\
& \text{subject to}
	& & F_{0}+\sum_{i=1}^{m} p_{i} F_{i}+X=0 \\
 & & & X \succeq 0. 
\end{aligned}
\label{prime}
\end{equation} 

We assume the primal and dual feasible sets defined as
\be
\ba{l}
\dps \mathcal{F}^{p} = \lc X | X=-F_{0}-\sum_{i=1}^{m} p_{i} F_{i}\succeq 0, p\in \R^{m} \rc, \\
\dps \mathcal{F}^{d} = \lc Z | \<F_i,Z\> + b_{i} =0, Z\succeq 0\rc 
\ea
\label{feasible} 
\ee
are not empty. 
Under this condition, for any primal-dual pair $(X,Z)$ that belongs to the feasible sets in (\ref{feasible}),
the primal cost $\<b,p\>$ is always greater than or equal to the dual cost $\<F_0,Z\>$. 
The difference between the primal and dual costs for a feasible pair $(X,Z)$ is called the \emph{duality gap}. 
The duality gap is a measure of the optimality of a primal-dual pair.
The smaller the duality gap, the more optimal the solution pair $(X,Z)$ is. 
For (\ref{prime}) and (\ref{dual}), the duality gap is the function
\be
\dps G(X,Z)=\Tr{XZ}. 
\label{dualgap}
\ee
Indeed, 
\begin{eqnarray*}
\Tr{XZ}&=& \Tr{\l -F_{0}-\sum_{i=1}^{m} p_{i} F_{i} \r Z} = -\Tr{F_0 Z} - \sum_{i=1}^{m} p_{i} \Tr{F_i Z} \\
 &=&\<b,p\> - \<F_0,Z\>.  
\end{eqnarray*}
Finally, if we assume that both problems are strictly feasible i.e. the
sets
\be
\ba{l}
\dps \mathcal{F}^{p'} = \lc X | X=-F_{0}-\sum_{i=1}^{m} p_{i} F_{i}\succ 0, p\in \R^{m} \rc, \\
\dps \mathcal{F}^{d'} = \lc Z | \<F_i,Z\> + b_{i} =0, Z\succ 0\rc 
\ea
\label{sfeasible} 
\ee
are not empty, then there exists an optimal primal-dual pair $(X^{*},Z^{*})$ such that 
\be
\dps \Tr{X^{*}Z^{*}}=0. 
\label{optimal}
\ee 
Moreover, the primal and dual optimal costs are guaranteed to be finite. 
The condition in (\ref{optimal}) implies that for strictly feasible problems, 
the primal and dual costs are equal at their respective optimal points $X^{*}$ and $Z^{*}$.
Note that in the strictly feasible problem, the semi-definite constraints become definite constraints. 

The canonical way of dealing with constrained optimization is by first adding to the cost 
function a term that increases significantly if the constraints are not met, and then solve the
unconstrained problem by minimizing the new cost function. 
This technique is commonly referred to as the \emph{relaxation} of the constraints.
For example, lets assume that the problems in (\ref{prime}) and (\ref{dual}) are strictly feasible. 
The positive-definite constraints $X\succ 0$ and $Z\succ 0$, which defines the \emph{interior} of 
a pair of semi-definite cones, can 
be relaxed using an indicator function $I(X,Z)$ such that
\be
\dps I : \l X,Z \r \ra \lc \ba{lc} 0, & X\succ 0, Z\succ 0 \\
+\infty, & \textrm{otherwise} \ea \right.
\label{indicator}
\ee
The intuition behind relaxation using an indicator function is as follows. If the primal-dual 
pair $(X,Z)$ approaches the boundary of the interior region, then the indicator function 
$I(X,Z)$ approaches infinity, thus incurring a large penalty on the cost function.

The indicator function in (\ref{indicator}) is not useful for optimization because it is not differentiable. 
Instead, the indicator function can be replaced by a family of smooth, convex functions $B(X,Z)$ that not only approximate
the behavior of the indicator function but are also \emph{self-concordant}. 
We refer to these functions as \emph{barrier} functions. 
A scalar function $F: \R \ra \R $, is said to be self-concordant if it is at least three times differentiable and satisfies the inequality
\be
\dps | F'''(x) | \leq 2 F''(x)^{\frac{3}{2}}. 
\label{selfconc}
\ee
The concept of self-concordance has been generalized to vector and matrix functions, thus we can also
find such functions for the positive-definite variables $X$ and $Z$. 
Here we state, without proof, the key property of self-concordant functions.
\begin{property}
Functions that are self-concordant can be minimized in polynomial time to a given non-zero accuracy using a Newton type iteration~\cite{Nemirovskii89}. 
\end{property}
Examples of self-concordant functions include linear functions, quadratic functions, and logarithmic functions.
A valid barrier function for the semi-definite constraints from (\ref{prime}) and (\ref{dual}) is
\be
\dps B(X,Z)=-\log{\det{X}}-\log{\det{Z}}. 
\label{barrier}
\ee

\section{An Interior Point Algorithm and Its Properties}
We now describe an example primal-dual interior point algorithm. We focus on the key property of 
convergence. We show its usefulness in constructing the inductive invariants to be applied towards 
documenting the software implementation. 
The algorithm is displayed in table~\ref{alg:1} and is based on the work in~\cite{Monteiro97}. 

\subsection{Details of the Algorithm}
The algorithm in table~\ref{alg:1} is consisted of an initialization routine and a {\tt while} loop. 
The operator \annotcf{length} is used to compute the size of the input problem data. 
The operator \annotc{\^{}$^{-1}$} represents an algorithm such as QR decomposition that returns the
inverse of the matrix. 
The operator \annotc{\^{}$^{0.5}$} represents an algorithm such as Cholesky decomposition 
that computes the square root of the input matrix. 
The operator \annotcf{lsqr} represents a least-square QR factorization algorithm that is used to
solve linear systems of equation of the form $Ax=b$. 
With the assumption of real algebra, all of these operators return exact solutions. 

In the initialization part, the states $X$, $Z$ and $p$ are initialized to feasible values, 
and the input problem data are assigned to constants $F_{i}, i=1,\ldots,m$. 
The term feasible here means
that $X$, $Z$, and $p$ satisfies the equality constraints of the primal and dual problems. 
We discuss more about the efficiency of the initialization process later on. 

The {\tt while} loop is a Newton
iteration that computes the zero of the derivative of the potential function
\be
\dps \phi (X,Z)=\l n+ \nu \sqrt{n} \r \log{\Tr{XZ}} - \log{\det{XZ}} - n \log{n}, 
\label{potential2}
\ee
in which $\nu$ is a positive weighting factor. 
Note that the potential function is  
a weighted sum of the primal-dual cost gap and the barrier function potential. 
The weighting factor $\nu$ 
is used in computing the duality gap reduction factor
$\sigma \equiv \frac{n}{n +\nu \sqrt{n}}$. 
A larger $\nu$ implies a smaller $\sigma$, which then implies a shorter convergence
time. For our algorithm, since we use a fix-step size of $1$, a small enough 
$\sigma$ combined with the newton step could result in a pair of $X$ and $Z$ that 
no longer belong to the interior of the positive-semidefinite cone. 
In the running example, we have $\nu=0.4714$. While this choice of $\nu$ doubled 
the number of iterations of the running example compared to the typical choice of $\nu=1$,
however it is critical in satisfying 
the invariants of the {\tt while} loop that are introduced later this paper. 
\begin{table}[htp]
{\footnotesize
\begin{center}
\begin{tabular*}{\linewidth}{@{}llrr@{}}
{\bfseries Algorithm 1. MZ Short-Path Primal-Dual Interior Point Algorithm}\\
\hline {\bf Input:} $F_0 \succ 0$, $F_i\in \mathbb{S}^{n},i=1,\hdots,m$, $b\in \R^{m}$\\
~~$\qquad$ $\epsilon$: Optimality desired\\
\hline \\
~1. Initialize: \\
~~$\qquad$ Compute $Z$ such that $\<F_i,Z\>=-b_i,i=1,\ldots,m$; \\
~~$\qquad$ Let $X \gets \hat{X}$; \emph{// $\hat{X}$ is some positive-definite matrix} \\
~~$\qquad$ Compute $p$ such that $\sum_{i}^{m} p_{i} F_{i}=-X_0-F_0$; \\
~~$\qquad$ Let $\mu \gets \frac{\<X,Z\>}{n}$; \\ \\
~~$\qquad$ Let $\sigma\gets 0.75$;\\
~~$\qquad$ Let $n\gets \size{F_i}$, $m\gets \size{b_i}$; \\
~2. {\bf while} $n\mu > \epsilon$ \{\\
~3. $\qquad$ Let $\phi_{-} \gets \<X,Z\>$; \\
~4. $\qquad$ Let $T_{inv} \gets Z^{0.5}$; \\
~5. $\qquad$ Let $T \gets T_{inv}^{-1}$; \\
~6. $\qquad$ Compute $(\Delta Z, \Delta X, \Delta p)$ that satisfies (\ref{newton12}); \\
~7. $\qquad$ Let $Z\gets Z +  \Delta Z$, $X \gets X + \Delta X$, $p\gets p +  \Delta p$; \\
~8. $\qquad$ Let $\phi \gets \<X,Z\>$; \\
~9. $\qquad$ Let $\mu \gets \frac{\<X,Z\>}{n}$; \\ \\
~10. $\qquad$ {\bf if } ($\phi-\phi_{-}>0$) \{ \\
	11.$\qquad$ $\qquad$ $\return{}$; \\
~~~~$\qquad$ \}\\
~~~~\} \\
\hline
\end{tabular*}
\end{center}
\caption{Primal-Dual Short Path Interior Point Algorithm}
\label{alg:1}} 
\end{table}

Let symbol $T=Z^{-0.5}$ and $T_{inv}$ denotes the inverse of $T$. 
The {\tt while} loop solves the set of matrix equations
\be
\begin{aligned}
	\<F_i, \Delta Z \> & = 0 \\
	\sum_{i}^{m} \Delta p_{i} F_{i} + \Delta X &=0\\
	\frac{1}{2} \l T \l Z \Delta X + \Delta Z X \r T_{inv} + T_{inv} \l \Delta X Z + X \Delta Z \r T \r &= \sigma \mu I - T_{inv} X T_{inv}. 
\end{aligned}
\label{newton12}
\ee
for the Newton-search directions $\Delta Z$, $\Delta X$ and $\Delta p$. 
The first two equations in (\ref{newton12}) are obtained from a Taylor expansion of the 
equality constraints from the primal and dual problems. 
These two constraints formulates the feasibility sets as defined in Eq (\ref{feasible}). 
The last equation in (\ref{newton12}) is obtained by 
setting the Taylor expansion of the derivative of (\ref{potential2}) equal to $0$, 
and then applying the symmetrizing transformation
\be
\ba{lc}
	\dps H_{T}: M \ra \frac{1}{2} \l T M T^{-1} + \l TMT^{-1}\r^{\m{T}}\r, & T=Z^{-0.5}
\ea
\label{ht}
\ee
to the result. 
To see this, note that derivative of (\ref{potential2}) is 
$\lb \ba{cc} XZ - \frac{n}{n+\nu \sqrt{n}} \frac{\Tr{XZ}}{n} I &\ea \right.$ 
$ \left. \ba{c} ZX - \frac{n}{n+\nu \sqrt{n}} \frac{\Tr{XZ}}{n} I \ea \rb$. 

The transformation in (\ref{ht}) is necessary to guarantee the solution $\Delta X$ is symmetric. 
The parameter $\sigma$, as mentioned before, can be  interpreted as a duality gap reduction factor. 
To see this, note that the 3rd equation in (\ref{newton12}) is the result of applying Newton 
iteration to solve the equation $XZ = \sigma \mu I$. 
With $\sigma \in (0,1)$, the duality gap $\Tr{XZ}=n \sigma \mu$ is reduced after every iteration. 
The choice of $T$ in (\ref{ht}) is taken from~\cite{Monteiro98aunified} and is called the Monteiro-Zhang (MZ) direction. 
Many of the Newton search directions from the interior-point method literature can be derived from an appropriate choice of $T$. 
The M-Z direction also guarantees an unique solution $\Delta X$ to (\ref{newton12}). 
The {\tt while} loop then updates the states $X,Z,p$ with the computed search directions and computes the new normalized duality gap. 
The aforementioned steps are repeated until the duality gap $n \mu$ is less than the desired accuracy $\epsilon$. 

\subsection{High-level Functional Property of the Algorithm} 
The key high-level functional property of the interior point algorithm in \ref{alg:1} is an upper bound on the worst case 
computational time to reach the specified duality gap $\epsilon>0$. 
The convergence rate is derived from a constant reduction in the potential function in (\ref{potential2})~\cite{mcgovern_feron}
after each iteration of the {\tt while} loop. 

Given the potential function in (\ref{potential2}), the following result gives us a tight upper 
bound on the convergence time of our running example. 
\begin{theorem}
	Let $X_{-}$, $Z_{-}$, and $p_{-}$ denote the values of $X$, $Z$, and $p$ in the previous iteration. 
	If there exist a constant $\delta>0$ such that
	\be
	\dps  \phi(X_{-},Z_{-})-\phi(X,Z) \geq \delta, 
	\label{reduction}
	\ee
	then Algorithm \ref{alg:1} will take at most
	$\mathcal{O} \l \sqrt{n} \log{\epsilon^{-1} \Tr {X_{0} Z_{0}}} \r$ iterations to converge to a duality gap of $\epsilon$, 	
	\label{reduction1}
\end{theorem}
For safety-critical applications, it is important for the optimization program implementation to have a 
rigorous guarantee of convergence within a specified time. 
Assuming that the required precision $\epsilon$ and the problem data size $n$ are known a priori, we can guarantee 
a tight upper bound on the optimization algorithm if the function $\phi$ satisfies (\ref{reduction}). 
For the running example, this is indeed true.  We have the following result. 
\begin{theorem}
	There exists a constant $\delta>0$ such that theorem \ref{reduction1} holds.
	\label{reduction2}
\end{theorem}
The proof of theorem \ref{reduction2} is not shown here for the sake of brevity but it is based on proofs already 
available in the interior point method literature (see~\cite{Kojima97} and~\cite{Nesterov95primal}).

Using theorems \ref{reduction1} and \ref{reduction2}, we can conclude that the algorithm in table~\ref{alg:1}, 
at worst, converges to the $\epsilon$-optimal solution linearly. 
For documenting the {\tt while} loop portion of the 
implementation, however we need to construct an inductive invariant of the form
\be
\dps 0\leq \phi(X,Z) \leq c,
\label{invariant:main} 
\ee
in which $c$ is a positive scalar. 
While the potential function in (\ref{potential2}) is useful for the construction of the algorithm in table~\ref{alg:1},
but it is not non-negative. 
To construct an invariant in the form of (\ref{invariant:main}), 
instead of using (\ref{potential2}), consider 
\be
\dps \phi(X,Z)=\log{\Tr{XZ}},  
\label{potential3}
\ee
which is simply the log of the duality gap function. 
\begin{theorem}
\label{reduction4}
The function in (\ref{potential3}) satisfies theorems \ref{reduction1} and \ref{reduction2}. 
\end{theorem}
An immediate implication of theorem \ref{reduction4} is that $\Tr{XZ}$ converges to $0$ linearly i.e. 
$\dps \exists \kappa\in (0,1)$ such that $\Tr{XZ} \leq \kappa \Tr{X_{-}Z_{-}}$, in which $X_{-}$ and
$Z_{-}$ are values of $X$ and $Z$ at the previous iteration. 
Using $\Tr{XZ}$, we can construct the invariant from 
(\ref{invariant:main}) and another invariant $\phi - \kappa \phi_{-} < 0$
to express the convergence property from theorem \ref{reduction1}. 

Additionally, there is one other invariant to be documented for the {\tt while} loop. 
The first one is the positive-definiteness of the states $X$ and $Z$. 
We need to show that the initial $X$ and $Z$ belongs to a positive-definite cone. 
We also need to show that they are guaranteed to remain in that cone throughout the execution of the {\tt while} loop. 
This inductive property is directly obtained from the constraints on the variables $X$ and $Z$. 
It is also used to show that the duality gap $\Tr{XZ}$ is positive. 

\section{Running Example} 
In this section, we introduce a basic optimization problem and a matlab program that solves the problem using the 
interior-point algorithm in table~\ref{alg:1}. 
\subsection{Input Problem}

The input data is obtained from a generic optimization problem taken from systems and control. 
The details of the original problem is skipped here as it has no bearing on the 
main contribution of this article. We do like to mention that 
the matrices $F_i, i=0,\ldots,3$ are computed from the original problem using the tool Yalmip~\cite{YALMIP}. 

\subsection{Matlab Implementation}
The Matlab implementation contains some minor differences from the algorithm description. 
The first is that, in the Matlab implementation, the current values of $X$, $Z$, $p$ are assigned to
the variables $X_{-}$, $Z_{-}$, and $p_{-}$ at the beginning of the {\tt while} loop. 
Note that the variables $X_{-}$, $Z_{-}$, and $p_{-}$ 
are denoted by \annotc{Xm}, \annotc{Zm} and \annotc{pm} respectively in the 
Matlab code. 
Because of that, steps 3 to 6 of algorithm in Table \ref{alg:1} are executed with the variables 
$X_{-}$, $Z_{-}$, and $p_{-}$ instead of $X$, $Z$, and $p$. 
Accordingly, step 7 becomes $Z\gets Z_{-} +  \Delta Z_{-}$, $X \gets X_{-} + \Delta X_{-}$, $p\gets p_{-} +  \Delta p_{-}$. 
This difference is the result of the need
to have the invariant of the form $\phi\l X,Z\r \leq \kappa \phi \l X_{-}, Z_{-}\r$ for $\kappa\in (0,1)$, which is important since
it expresses the fast convergence property from theorem \ref{reduction1}. 

\section{Annotating an Optimization Program for Deductive Verification}

In this section, we discuss, line by line, 
an example of a fully-annotated
optimization program that is based on the interior-point algorithm in figure~\ref{alg:1}. 
The fully-annotated program represents a claim of proof that the code also conforms to
certain safety or liveness property of the algorithm. 
The claim of proof need to be verified 
by a proof-checking program, which we called the {\tt backend}. 
The annotated program was written in the Matlab language and the Hoare logic 
style annotations were inserted manually. 
Matlab was chosen as the implementation language because of its 
compactness and readability. 
For the automation of the process i.e. the credible autocoding framework, 
we shall choose a realistic target language such as C that is 
more acceptable for formal analysis and verification. 

The annotation process starts with the selection of 
a safety or liveness property of the interior-point algorithm in which 
the code originates from.
The chosen property is formalized and then translated into invariants and then
inserted into the code as Hoare logic pre/post conditions. 
For the example optimization program, the invariant that encode 
the property of bounded execution time of the program is
$\phi - 0.76\phi_{-} < 0$. 
Various additional proof elements are inserted into the
code, also in the form of Hoare logic style pre/post-conditions, 
to allow easy automation of the proof-checking process. 
The amount of additional proof elements needed is dependent on
the capability of the backend. 
Lastly, by using the Hoare logic rules, 
the entire program is populate with pre/post-conditions, 
thus resulting in the fully-annotated code. 

The rest of this section is organized as follows. First we give the example Matlab implementation and 
a description of the non-standard Matlab functions used in the code. Second, 
we give a description of the annotation language used in expressing the invariants and assertions.
Finally, we give a detailed description of all the Hoare triples in the fully-annotated example.

\subsection{Matlab Implementation} 
The example implementation uses three non-standard Matlab operators 
\annotc{vecs}, \annotc{mats} and \annotc{krons}. 
These operators are used to transform
the matrix equations in(\ref{newton12}) into matrix vector equations in the form $Ax=b$. 
For more details about these operator, please refer to the appendix. 
The symbol \annotc{'} denotes Matlab's transpose function. 

\begin{figure}
\begin{lstlisting}
F0=[1, 0; 0, 0.1];
F1=[-0.750999 0.00499; 0.00499 0.0001];
F2=[0.03992 -0.999101; -0.999101 0.00002];
F3=[0.0016 0.00004; 0.00004 -0.999999];
b=[0.4; -0.2; 0.2];
n=length(F0);
m=length(b);
F=[vecs(F1); vecs(F2); vecs(F3)];
Ft=F';
Z=mats(lsqr(F,-b),n); 
X=[0.3409 0.2407; 0.2407 0.9021];
epsilon=1e-8;
sigma=0.75;
phi=trace(X*Z); 
phim=1/0.75*phi; 
P=mats(lsqr(Ft,vecs(-X-F0)),n);
p=vecs(P);
while (phi>epsilon) {
    Xm=X;
    Zm=Z;
    pm=p;
    mu=trace(Xm*Zm)/n;
    Zh=Zm^(0.5); 
    Zhi=Zh^(-1); 
    G=krons(Zhi,Zh'*Xm,n,m);
    H=krons(Zhi*Zm,Zh',n,m);
    r=sigma*mu*eye(n,n)-Zh*Xm*Zh;
    dZm=lsqr(F,zeros(m,1));
    dXm=lsqr(H, vecs(r)-G*dZm);
    dpm=lsqr(Ft,-dXm); 
    p=pm+dpm
    X=Xm+mats(dXm,n);
    Z=Zm+mats(dZm,n); 
    phim=trace(Xm*Zm);
    phi=trace(X*Z); 
    mu=trace(X*Z)/n;
}
\end{lstlisting} 
\caption{Optimization Program in Matlab} 
\label{uannot} 
\end{figure}

The program is divided into two parts. The first part, which we call the initialization part,
assigns the data of the input optimization problem to memory and initializes the variables to
be optimized. 
The second part, which is the {\tt while} loop, executes the interior-point algorithm to solve
the input optimization problem. 

\subsection{Annotation Language} 

The fully-annotated example is displayed in section~\ref{sec:annot} of the appendix. 
The annotations are expressed using a pseudo Matlab specification language (MSL) that is similar, in features, 
to the ANSI/ISO C Specification Language (ACSL) for C programs~\cite{Cuoq:2012}, 
Like in ACSL, the keywords \annotf{requires} and \annotf{ensures} denote a pre and post-condition statement respectively.
A MSL contract, like the ACSL contract, is used to express a Hoare triple. 
For example, the MSL contract displayed in figure~\ref{contract} can be parsed
as the Hoare triple $\lc P \rc C; \lc Q \rc$. 
\begin{figure}
\begin{lstlisting}
% requires P
% ensures  Q
{ 
  C;
}
\end{lstlisting} 
\caption{An ACSL-like Contract}
\label{contract}
\end{figure}

A block of code can have more than one contract. 
For example, as shown in figure~\ref{multiple}, 
the two MSL contracts translates to the Hoare triples 
\be
\dps \lc P_i \rc C; \lc Q_i \rc, i=1,2. 
\label{contracts}
\ee
\begin{figure}
\begin{lstlisting}
% requires P1
% ensures  Q2

% requires P2
% ensures  Q2
{ 
  C;
}
\end{lstlisting} 
\caption{Multiple Contracts} 
\label{multiple} 
\end{figure}

A block of empty code can also have contract.
For example, the formula $P\Rightarrow Q$ can be
expanded into a Hoare triple $\lc P \rc empty;\lc Q \rc$,
which is expressed using the MSL contract in figure~\ref{uannot:empty}. 
\begin{figure}
\begin{lstlisting}
% requires P
% ensures  Q
{ 
  % empty;
}
\end{lstlisting} 
\caption{Contract for an empty code} 
\label{uannot:empty}
\end{figure}

\begin{remark} 
The block of empty code appear in our annotated program, 
for example when we expand the post-condition of Hoare triple 
such as
\be
\lc P \rc C; \lc Q_1 \Rightarrow Q_2 \rc
\label{one}
\ee
into an additional Hoare triple.
This transformation results in
(\ref{one}) being expanded into two Hoare triples
\be
\lc P \rc C; \lc Q_1 \rc
\label{two}
\ee and 
\be
\lc Q_1 \rc empty; \lc Q_2 \rc. 
\label{three}
\ee
The main reason to expand a Hoare triple into multiple ones
is to simplify the automation of the proof-checking process.
The choice in expanding is arbitrary as it is dependent on the
capability of the backend.
For example, in the case of (\ref{one}), it is possible that
the backend can verify (\ref{two}) and (\ref{three}) separately
but not (\ref{one}) without human input. 
Practically speaking, by reducing the complexity in the verification of an
individual Hoare triple in exchange for an increase 
in the total number of Hoare triples, 
the proof-checking process usually become easier to automate. 
\end{remark} 

A contract can have more than one pre-condition
statement as shown in figure~\ref{uannot:mult}. 
The pre-condition statements are combined into a 
single pre-condition $P$ conjunctively when the contract
is parsed as a Hoare triple. 
For example, the contract in figure~\ref{uannot:mult}
expresses the Hoare triple
$\lc P_1 \wedge P_2 \rc empty; \lc Q \rc$. 
\begin{figure}
\begin{lstlisting}
% requires P_1
% requires P_2
% ensures  Q
{ 
  C;
}
\end{lstlisting} 
\caption{Contract for an empty code} 
\label{uannot:mult}
\end{figure}

The logic comparison operator
$>$ is overloaded to express $\succ$. 
The function \annotf{smat} is the inverse of the function \annotf{svec}. 
The function \annotf{svec} is similar to the symmetric vectorization
function \annotf{vecs} but with a multiplication factor of $2$. 
For example, the expression \annotf{smat([0.4;-0.2;0.2]} returns the matrix
\be
\dps \begin{bmatrix} 0.4 & -0.1 \cr -0.1 & 0.2 \end{bmatrix}. 
\label{smat1}
\ee

\subsection{Functions with Verified Contracts} 

Certain functions used in the example program comes
by default with MSL contracts to
ensures the regularity of the inputs and outputs to the functions. 
These annotations are implicit since they are assumed to be correct. 

For example, the square root function and the inverse function 
denoted by the symbols \annotc{\^{0.5}} and \annotc{\^{-1.0}} respectively, 
comes by default with the properties 
that the input matrix is symmetric positive-definite and the output is
also symmetric positive-definite. 
\begin{lstlisting}
% requires Zm>0;
% ensures  Zh>0;
Zh=Zm^(0.5);
% requires Zh>0;
% ensures  Zhi>0;
Zhi=Zh^(-1)
\end{lstlisting}
The function \annotc{lsqr} comes by default with the post-condition which stipulates that
the matrix-vector product of the first argument of the function and the output of the function is 
equal to the second argument of the function. 
\begin{lstlisting}
% ensures F*y==-b;
y=lsqr(F,-b);
\end{lstlisting} 
The function \annotc{vecs} by default comes with the pre-condition that the input argument to the
function is a symmetric matrix. 
\begin{lstlisting}
% requires P==transpose(P); 
p=vecs(P); 
\end{lstlisting} 
The function \annotc{mats}, as displayed below, requires that $x\in \R^{\frac{n\l n+1\r}{2}}$ and
ensures that the output $X$ is a symmetric matrix of size $n$. 
\begin{lstlisting}
% requires n>=1; 
% requires type(x)==vector(n/2*(n+1));
% ensures  type(X)==symmetric_matrix(n);
X=mats(x,n);
\end{lstlisting}

\subsection{Annotated Code} 
 
For the analysis and discussion of the annotations, 
the fully-annotated example is split into multiple parts and each 
part is displayed in a separate figure and discussed in a separate 
subsection. 
In those figures, every MSL contract is assigned 
a pair of numbers $(n,m)$, 
which indicates the $m$th contract of 
the $n$th block of code of the figure. 
The Hoare triple expressed 
by the contract with a label of $(n,m)$ is denoted
as
\be
\dps \mathcal{H}_{n,m}:=\lc P_{n,m} \rc C_{n}; \lc Q_{n,m}\rc. 
\label{hoare_triple}
\ee
For example, in figure~\ref{hoare_example}, 
$\mathcal{H}_{1,1}$ is the Hoare triple $\lc x>0\rc x=y; \lc y>0\rc$, 
with $P_{1,1}$ being the pre-condition $x>z$, 
$C_{1}$ being the line of code \annotc{y=x-z;} 
and $Q_{1,1}$ being the post-condition $y>0$. 
\begin{figure}
\begin{lstlisting}
% (1,1) requires  x>z; 
%       ensures   y>0; 

% (1,2) requires  x<0;
%       requires  z>0;
%       ensures   y<0; 
{
  y=x-z;
}
\end{lstlisting}
\label{hoare_example}
\end{figure}
Also in figure~\ref{hoare_example}, $\mathcal{H}_{1,2}$ is the
Hoare triple $\lc x<0 \wedge z>0\rc y=x-z; \lc y<0 \rc$, 
with $P_{1,2}:=x<0 \wedge z>0$, and $Q_{1,2}:=y<0$. 

The variables from the program will be referenced 
using either their programmatic textual representation or their corresponding 
mathematical symbols. 
For example, the list of variables \annotc{phi}, \annotc{phim}, \annotc{dXm}, and \annotc{Xm } is the same as
the list of symbols $\phi$, $\phi_{-}$, $\Delta_{X_{-}}$, and $X_{-}$ 
Likewise, the annotations are also referenced using both representations as well. 
For example, the expression \annotc{phim=trace(Xm*Zm)/n} is equivalent to $\phi_{-}:= \frac{\Tr{X_{-} Z_{-}}}{n}$. 
Some expressions from the program such as \annotc{mats(dXm,n)} and \annotc{mats(dZm,n)} are referenced using the
more compact symbols $\Delta X_{-}$ and $\Delta Z_{-}$ respectively. 

\subsection{Proof Checking the Annotations} 

We assume there exists an automated proof-checking program,  which we referred to as
the {\tt backend}, that can 
be used to verify the Hoare triples in the annotated example.  
We will not go into much details about the backend other than occasionally 
discussing some of theories and formulas needed to verify certain annotations. 

\subsection{Initialization Part I}

The first part of initialization code along with its annotations 
is displayed in figure~\ref{annot:init1}. 
The annotations described 
in this subsection of the paper are from that figure unless explicitly 
states otherwise. 
\begin{figure}
\begin{lstlisting}
% (1,1) ensures F0>0; 
{
  F0=[1, 0; 0, 0.1];
}
% (2,1) ensures transpose(F1)==F1; 
{ 
  F1=[-0.750999 0.00499; 0.00499 0.0001];
}
% (3,1) ensures transpose(F2)==F2;
{
  F2=[0.03992 -0.999101; -0.999101 0.00002];
}
% (4,1) ensures transpose(F3)==F3;
{
  F3=[0.0016 0.00004; 0.00004 -0.999999];
}
% (5,1) ensures smat(b)>0; 
{
  b=[0.4; -0.2; 0.2];
}
F=[vecs(F1); vecs(F2); vecs(F3)];
% (6,1) ensures Ft==transpose(F);
{
  Ft=F'; 
}
% (7,1) ensures n>=1;
{ 
  n=length(F0);
}
% (8,1) ensures m>=1;
{
  m=length(b);
}
\end{lstlisting}
\caption{Initialization}
\label{annot:init1}
\end{figure}

In $C_{i},i=1,\ldots,5$,
the data parameters of the input optimization problem are assigned
to the appropriate variables.
The inserted post-conditions $P_{i,1}, i=1,\ldots,5$
represent regularity conditions on the data parameters of the input problem.
The correctness of these post-conditions guarantee that the input optimization problem is well-posed. 
For example, the post-condition $P_{1,1}$ claims that 
the variable \annotc{F0} is positive-definite after the execution of $C_{1}$ in line 2. 
Another example is $Q_{2,1}$ which claims that after execution of $C_{2}$ in line 4,
the variable \annotc{F1} is a symmetric matrix. 

In $C_{6}$ and $C_{y}$, the problem sizes are computed. 
In our example, the variable \annotc{m} 
denotes the number of equality constraints in the dual formulation and
the variable \annotc{n}
denotes the dimensions of the optimization variables $X$ and $Z$. 
The inserted post-conditions $Q_{6,1}$ and $Q_{7,1}$ checks
that the problem sizes are at least one.

\subsection{Intialization Part II}
The annotations described in this subsection of the paper are
from figure~\ref{annot:init2} unless explicitly 
states otherwise. 
\begin{figure}
\begin{lstlisting}
% (1,1) ensures Z>0; 
{
  Z=mats(lsqr(F,-b),n); 
}
% (2,1) ensures X>0 

% (2,2) ensures trace(X*Z)<=0.1; 
{
  X=[0.3409 0.2407; 0.2407 0.9021];
}
% (3,1) requires Z>0 && X>0; 
%       ensures  trace(X*Z)>0;
{
  % empty code
}
% (4,1) ensures transpose(P)==P; 
{
  P=mats(lsqr(Ft,vecs(-X-F0)),n);
  p=vecs(P);
}
% (5,1) ensures epsilon>0
{
  epsilon=1e-8;
}
% (6,1) ensures sigma==0.75;
{
  sigma=0.75;
}
% (7,1) requires trace(X*Z)<=0.1;
%       ensures  phi<=0.1; 

% (7,2) requires trace(X*Z)>0; 
%       ensures  phi>0; 

% (7,3) ensures  phi==trace(X*Z); 
{
  phi=trace(X*Z); 
}
% (8,1) requires phi>0;
%       ensures  phi-0.76/0.75*phi<0; 
{
  % empty code
}
% (9,1) requires phi-0.76/0.75*phi<0; 
%       ensures  phi-0.76*phim<0; 
{
  phim=1/0.75*phi; 
}
% (10,2) ensures  mu==trace(X*Z)/n;
{
  mu=trace(X*Z)/n;
}
\end{lstlisting}
\caption{Initialization Part II}
\label{annot:init2}
\end{figure}

The blocks of code $C_{i}, i=1,2$ generate and then assign 
initial values to variables $X$ and $Z$. 
The values of variables $X$ and $Z$ need to be positive-definite
in order to satisfy the 
strict feasibility property of the input optimization problem. 
Furthermore, 
functions used in the latter part of the program 
such as \annotc{\^{0.5}} and \annotc{\^{-1}} implicitly 
require the input to be positive-definite. 
To check if $X$ and $Z$ are positive-definite after they are initialized, 
we insert the post-conditions $Q_{1,1}$ and $Q_{2,1}$. 
The correctness of $\mathcal{H}_{i,1},i=1,2$ can be checked
by a backend that can verify if a matrix is positive
definite. 
For  $C_{2}$, we insert a second contract to check if 
the quantity of the duality gap, defined as the
inner product $X$ and $Z$ or \annotc{trace(X*Z)}, is bounded from above by $0.1$. 

Using the skip rule from~\ref{hoare_rules},
the post-conditions $Q_{1,1}$ and $Q_{2,1}$
are propagated forward, and combined conjunctively 
to form $P_{3,1}$. 
The pre-condition $P_{3,1}$ is an invariant to be annotated for the {\tt while} loop
By the application of the theory
\be
\dps X \succ 0 \wedge Z\succ0 \Rightarrow \Tr{XZ}>0,
\label{strat1}
\ee
we get the post-condition $Q_{3,1}$. 
The post-condition $Q_{3,1}$ is a claim that the duality gap is also bounded 
from below by $0$. 
For the automatic checking of $\mathcal{H}_{3,1}$, the back-end can 
use the same theory from (\ref{strat1}). 

Next in $C_{4}$, the initial value for $p$ is computed using the matrix equality 
$\dps F0+\sum_{i}^{m} p_{i} F_{i}+X=0$, which is taken from the primal 
formulation. 
Here, we only insert the post-condition \annotf{transpose(P)==P}
since it is one of the implicit 
pre-condition of the function \annotf{vecs} in line 19. 

We move on to $C_5$, which assigns the value $1\times 10^{-8}$ 
to the variable \annotc{epsilon}. 
This value is essentially a measure of the desired optimality and it 
is also an important factor in computing an upper bound on the number of 
iterations of the {\tt while} loop. 
The proof of the bounded time termination of the loop requires
the desired optimality to be greater 
than $0$, hence the insertion of $Q_{5,1}$. 
The next block of code, which is $C_{6}$ assigns the value of 
$0.75$ to the variable $\sigma$. 
The inserted post-condition, which is $P_{6,1}$ ensures that the 
variable $\sigma$ is equal to the value $0.75$. 
This trivial condition is to be used later in the annotation process. 
The verification of $Q_{i,1},i=5,6$ should be automatic 
for any theorem provers. 

Next, we move ahead to $C_{7}$, which computes the duality gap 
$\Tr{X*Z}$ and then assigns the result to the variable \annotc{phi}.
There are three contracts for $C_{7}$. 
For $\mathcal{H}_{7,j},j=1,2$, the pre-conditions $P_{7,1}$ and $P_{7,2}$, 
which correspond to \annotf{Trace(X*Z)>0} and \annotf{Trace(X*Z)<=0.1} respectively, 
are obtained by applying the skip rule to $Q_{2,2}$ and $Q_{3,1}$. 
The post-conditions $Q_{7,1}$ and $Q_{7,2}$ are
generated using the backward substitution rule from~\ref{hoare_rules}.
They are also invariants to be annotated for the {\tt while} loop. 
The last contract of $C_{7}$ contains a trivially true post-condition 
\annotf{phi==trace(X*Z)} to be used later in the annotation process. 

Now moving on to $C_{8}$, we see another inserted 
empty block of code (line 42) with the pre-condition
\annotf{phi>0}. 
The pre-condition is simply $Q_{7,2}$ propagated forward. 
Before we discuss the post-condition in $\mathcal{H}_{8,1}$, 
we first jump ahead to next
block of code which is $C_{9}$. 
For $C_{9}$, the inserted post-condition \annotf{phi-0.76*phim<0} is
an invariant obtained by formalizing the property of finite termination of
the interior point algorithm. 
This post-condition is an invariant to be annotated for the {\tt while} loop. 
Applying the backwards substitution rule on $Q_{9,1}$, we get
the pre-condition $P_{9,1}$ which is exactly the post-condition
for $C_{8}$. 
Since $C_{8}$ is an empty block of code, the correctness of the 
Hoare triple $\mathcal{H}_{8,1}$ reduces to
verifying the formula $P_{8,1}\Rightarrow Q_{8,1}$ which is
equivalent to  showing 
\be
\dps  c>1 \wedge \phi>0 \Rightarrow \phi - c \phi <0 \wedge 0.76/0.75>1
\label{strat2}
\ee
which can be discharged automatically by existing tools such as certain 
SMT solvers~\cite{z3_smt}. 

Finally, we insert a trivially true post-condition for the last 
block of code in figure~\ref{annot:init2}. The post-condition simply
ensures that the variable \annotc{mu} is equal to the expression 
\annotc{trace{X*Z}/n} after the latter has been assigned to the former.

In the next few sections, we show that $Q_{7,1}\wedge Q_{7,2} \wedge  Q_{9,1} $
holds throughout execution of the {\tt while} loop, thereby proving that optimization
program terminates with an answer within a bounded time. 
We also show that $P_{3,1}$ hold
as well throughout the entire execution of the loop, thereby
completing the proof of $Q_{7,1}\wedge Q_{7,2} \wedge  Q_{9,1}$.

\subsection{The {\tt while} loop} 

The annotations and code discussed in this subsection are from
figure~\ref{annot:loop0} unless explicitly stated otherwise. 
\begin{figure}
\begin{lstlisting}
% (1,1) requires phi>0 && phi<=0.1;
%       ensures  phi>0 && phi<=0.1;

% (1,2) requires phi-0.76*phim<0; 
%       ensures  phi-0.76*phim<0; 


% (1,3) requires Z>0 && X>0;
%       ensures  Z>0 && X>0;
{
  while (phi>epsilon) do
	.
	.
	.
  end
}
\end{lstlisting}
\caption{The While Loop} 
\label{annot:loop0}
\end{figure}

The contracts on the {\tt while} loop are constructed using
the invariants $Q_{7,1}\wedge Q_{7,2}$, $Q_{9,1}$ 
and $P_{3,1}$, in which $P_{3,1}$, $Q_{7,1}$, $Q_{7,2}$ and $Q_{9,1}$ 
are from figure~\ref{annot:init1}. 
By the application of the while axiom from (\ref{while}), 
each invariant becomes the pre and post-condition of a contract.  
In the next few subsections of the paper, we discuss like before, the line by line
Hoare logic style proof of the correctness of $Q_{1,j},j=1,3$.
For partial correctness, we just need to show that $Q_{1,j},j=1,3$ hold 
before the execution of the loop and after every execution of the loop body. 
The total correctness comes from the finite termination property 
encoded by the invariant \annotf{phi-0.76*phim<0}. 

\subsection{Proving $\phi-0.76 \phi_{-}<0$ on the code: Part I} 

The annotations and code discussed in this subsection are from 
figure~\ref{annot:loop1} unless stated otherwise. 
\begin{figure}
\begin{lstlisting}
% (1,1) requires phi>0 && phi<=0.1 && phi==trace(X*Z); 
%       ensures  trace(X*Z)>0 && trace(X*Z)<=0.1; 
{
  % empty code
}
% (2,1) requires trace(X*Z)>0 && trace(X*Z)<=0.1; 
%       ensures  trace(Xm*Zm)>0 && trace(Xm*Zm)<=0.1; 
{
  Xm=X;
  Zm=Z;
}
  pm=p;
% (3,1) requires n>=1; 
%       ensures  n*mu==trace(Xm*Zm);
{
  mu=trace(Xm*Zm)/n;
}
\end{lstlisting}
\caption{{\tt while} Loop Body: Part I}
\label{annot:loop1}
\end{figure}

The annotation of the loop body starts with the insertion of the block of empty code in line 4.  
Using the skip rule on the invariant $Q_{1,1}$ from figure~\ref{annot:loop0} and 
$Q_{7,3}$ from figure~\ref{annot:init1}, 
we get the pre-condition $P_{1,1}$. 
Apply the theory
\be
\dps P \wedge expr1=expr2 \Rightarrow P[expr1/expr2] \vee P[expr2/expr1]
\label{strat0}
\ee
to $P_{1,1}$, we get the post-condition $Q_{1,1}$. 
The formula in (\ref{strat0}) is used for several more annotations in the 
loop body. In the case of going from $P_{1,1}$ to $Q_{1,1}$, we
substituted all instances of the expression \annotf{phi}
in $\phi>0 \wedge \phi\leq 0.1$ with the expression \annotf{trace(X*Z)}.
The verification of $\mathcal{H}_{1,1}$ is fairly straightforward in 
a theorem prover such as PVS. 

The next block of code, which is $C_{2}$, assigns
the value of $X$ and $Z$ to the variables $X_{-}$ and $Z_{-}$ respectively. 
The pre-condition for $C_{2}$ is $Q_{1,1}$ because
the post-condition of a prior contract is also a pre-condition of
the current contract. 
The post-condition $\Tr{X_{-} Z_{-}} >0 \wedge \Tr{X_{-}Z_{-}} \leq 0.1$ is
generated by the application of the backward substitution rule. 

Finally, the last contract in figure~\ref{annot:loop1}
is for $C_{3}$, which assigns the expression \annotc{trace(Xm*Zm)/n} to the
variable \annotc{mu}. 
The pre-condition $P_{3,1}$ is 
$Q_{7,1}$ from figure~\ref{annot:init1} propagated
forward using the skip rule. 
The post-condition $n\mu=\Tr{X_{-}Z_{-}}$ is generated using 
the formula
\be
\dps n>=1 \Rightarrow n \neq 0  \wedge n\neq 0 \wedge x=\frac{y}{n} \Rightarrow nx=y. 
\label{strat4}
\ee
The correctness of (\ref{strat4}) can be discharged automatically 
by a theorem prover such as PVS or most SMT solvers. 

\subsection{Proving $\phi-0.76 \phi_{-} <0$ on the code: Part II} 
The annotations and code described in this subsection are from figure~\ref{annot:loop2} unless
stated otherwise. 
\begin{figure}[htp]
\begin{lstlisting}
% (1,1) requires n*mu==trace(Xm*Zm); 
%       requires trace(Xm*mats(lsqr(F,zeros(m,1)),n))+trace(mats(lsqr(krons((Zm^(0.5))^(-1)*Zm,(Zm^(0.5))',n,m),vecs(sigma*mu*eye(n,n)-(Zm^(0.5))*Xm*(Zm^(0.5)))-krons((Zm^(0.5))^(-1),(Zm^(0.5))'*Xm,n,m)*lsqr(F,zeros(m,1))),n)*Zm)+trace(Xm*Zm)-sigma*n*mu==0; 
%       ensures  trace(Xm*mats(lsqr(F,zeros(m,1)),n))+trace(mats(lsqr(krons((Zm^(0.5))^(-1)*Zm,(Zm^(0.5))',n,m),vecs(sigma*mu*eye(n,n)-(Zm^(0.5))*Xm*(Zm^(0.5)))-krons((Zm^(0.5))^(-1),(Zm^(0.5))'*Xm,n,m)*lsqr(F,zeros(m,1))),n)*Zm)+trace(Xm*Zm)-sigma*trace(Xm*Zm)==0;
{
  % empty
}
% (2,1) requires trace(Xm*mats(lsqr(F,zeros(m,1)),n))+trace(mats(lsqr(krons((Zm^(0.5))^(-1)*Zm,(Zm^(0.5))',n,m),vecs(sigma*mu*eye(n,n)-(Zm^(0.5))*Xm*(Zm^(0.5)))-krons((Zm^(0.5))^(-1),(Zm^(0.5))'*Xm,n,m)*lsqr(F,zeros(m,1))),n)*Zm)+trace(Xm*Zm)-sigma*trace(Xm*Zm)==0; 
%       ensures  trace(Xm*mats(lsqr(F,zeros(m,1)),n))+trace(mats(lsqr(H,vecs(sigma*mu*eye(n,n)-Zh*Xm*Zh)-G*lsqr(F,zeros(m,1))),n)*Zm)+trace(Xm*Zm)-sigma*trace(Xm*Zm)==0; 
{
  Zh=Zm^(0.5); 
  Zhi=Zh^(-1); 
  G=krons(Zhi,Zh'*Xm,n,m);
  H=krons(Zhi*Zm,Zh',n,m);
}
% (3,1) requires trace(Xm*mats(lsqr(F,zeros(m,1)),n))+trace(mats(lsqr(H,vecs(sigma*mu*eye(n,n)-Zh*Xm*Zh)-G*lsqr(F,zeros(m,1))),n)*Zm)+trace(Xm*Zm)-sigma*trace(Xm*Zm)==0;
%       ensures  trace(Xm*mats(dZm,n))+trace(mats(dXm,n)*Zm)+trace(Xm*Zm)-sigma*trace(Xm*Zm)==0;

% (3,2) ensures  dZm==lsqr(F,zeros(m,1));
{
  r=sigma*mu*eye(n,n)-Zh*Xm*Zh;
  dZm=lsqr(F,zeros(m,1));
  dXm=lsqr(H, vecs(r)-G*dZm);
}
% (4,1) requires sigma==0.75; 
%       requires trace(Xm*mats(dZm,n))+trace(mats(dXm,n)*Zm)+trace(Xm*Zm)-sigma*trace(Xm*Zm)==0;
%       ensures  trace(Xm*mats(dZm,n))+trace(mats(dXm,n)*Zm)+trace(Xm*Zm)-0.75*trace(Xm*Zm)==0;
{ 
  % empty code
}
\end{lstlisting}
\caption{{\tt while} Loop Body: Part II} 
\label{annot:loop2}
\end{figure}

We look at the first contract,
which is for the inserted block of empty code in line 5. 
The first pre-condition statement \annotf{n*mu==trace(Xm*Zm)} 
is obtained by applying the skip rule to 
$Q_{3,1}$ from figure~\ref{annot:loop1}. 
The second pre-condition statement is instantiation of the following
true statement: 
given $\Delta X$ and $\Delta Z$ that satisfy the equations
\be
\dps \sum_{i=1}^{m} \<F_{i}, \Delta Z\> = 0, 
\label{taut:0}
\ee
and 
\be
\begin{split}
\dps
& \frac{1}{2} \l Z^{0.5} \l \Delta X Z \r Z^{-0.5} + Z^{-0.5} \l Z \Delta X \r Z^{0.5} \r \\
& = \dps \frac{-1}{2} \l Z^{0.5} \l X \Delta Z \r Z^{-0.5} + Z^{-0.5} \l \Delta Z X \r Z^{0.5} \r \\
& + \sigma \mu I - Z^{0.5} X Z^{0.5}, 
\end{split}
\label{taut:1}
\ee
then $\forall X,Z\in \R^{n\times n}$, 
\be
\dps \Tr{X\Delta Z} + \Tr{\Delta X Z } + \Tr{XZ} - \sigma n \mu = 0.
\label{tauto:0}
\ee
The correctness of (\ref{tauto:0}) can be seen by noting that
\be
\begin{split} 
\Tr{\Delta X Z} & = \Tr{\frac{1}{2} \l Z^{0.5} \l \Delta X Z \r Z^{-0.5} + Z^{-0.5} \l Z \Delta X \r Z^{0.5} \r} \\
		  & = \Tr{\dps \frac{-1}{2} \l Z^{0.5} \l X \Delta Z \r Z^{-0.5} + Z^{-0.5} \l \Delta Z X \r Z^{0.5} \r}\\
                  & +  \Tr{\sigma \mu I_{n\times n}} - \Tr{Z^{0.5} X Z^{0.5}}\\
                  & =  -\Tr{X \Delta Z} + \sigma n \mu - \Tr{XZ}, 
\end{split}
\label{taut:2}
\ee
hence
\be
\begin{split} 
	\dps & \Tr{X\Delta Z} + \Tr{\Delta X Z } + \Tr{XZ} - \sigma n \mu \\
	\dps & = \Tr{X\Delta Z} - \Tr{X \Delta Z} + \sigma n \mu - \Tr{XZ} + \Tr{XZ} - \sigma n \mu \\
	\dps & = 0
\end{split}
\label{taut:3}
\ee
Now we look at the pre-condition statement from line 2.  
First, note that the subexpression \annotf{mats(lsqr(F,zeros(m,1))} from line 2
returns a $\Delta Z$ that satisfies (\ref{taut:0}). 
Second, note that the subexpression \annotf{mats(lsqr(krons((...,} \annotf{vecs(sigma...*Zm)} from line 2
returns a $\Delta X$ that satisfies, with $X$ replaced by $X_{-}$ and $Z$ replaced by $Z_{-}$, the
equation in (\ref{taut:1}). 
Finally, one can see that 
the pre-condition statement is generated by instantiating $X$ and $Z$ of (\ref{tauto:0}) 
with $X_{-}$ and $Z_{-}$. 

The post-condition for $\mathcal{H}_{1,1}$ is generated using the formula from (\ref{strat0}) 
i.e. by replacing all instances of the expression \annotc{n*mu} 
in the pre-condition statement from line 2 
with the expression \annotc{trace(Xm*Zm)}. 

The next contract is consisted of the pre-condition 
$P_{2,1}$, which is a duplication of $Q_{1,1}$, 
and the post-condition $Q_{2,1}$.
The post-condition $Q_{2,1}$ is generated by four successive
application of the backward substitution rule, one for each line of 
code in $C_{2}$. This process results in four successive 
Hoare triples, which are merged into $\mathcal{H}_{2,1}$ by the 
application of the composition rule from~\ref{compose}. 

The Hoare triple $\mathcal{H}_{3,1}$ is constructed in a similar
fashion as $\mathcal{H}_{2,1}$. 
For $C_{3}$, we insert an additional contract ensuring that
the variable \annotc{dZm} is equal to the expression 
\annotc{lsqr(F,zeros(m,1))}. This post-condition is correct because of 
the code in line 21 and will be used later in the annotation of the loop body. 

The last contract from figure~\ref{annot:loop2}, which forms $\mathcal{H}_{4,1}$ has two 
pre-condition statements. 
The first pre-condition is 
$Q_{6,1}$ from figure~\ref{annot:init1} propagated forward using the skip rule. 
The second pre-condition statement is simply the post-condition of the prior contract.  
The post-condition $Q_{4,1}$ is generated using the formula in (\ref{strat0}).
In this case, the formula results in all instances of the expression 
\annotf{sigma} in $Q_{3,1}$ being replaced by the expression \annotf{0.75}. 

\subsection{Proving $\phi-0.76 \phi_{-} <0$ on the code: Part III}
The annotations and code described in this subsection are from figure~\ref{annot:loop2} unless
stated otherwise. 
\begin{figure}
\begin{lstlisting}
% (1,1) requires  dZm==lsqr(F,zeros(m,1));
%       ensures   lsqr(Ft,-dXm)'*F*dZm==0;
{ 
  % empty code
}
% (2,1) requires lsqr(Ft,-dXm)*F*dZm==0;
%       requires transpose(F)==Ft; 
%       ensures  dot(Ft*lsqr(Ft,-dXm),dZm)==0;
{
  % empty code
}
% (3,1) requires dot(Ft*lsqr(Ft,-dXm),dZm)==0;  
%       ensures  trace(mats(Ft*lsqr(Ft,-dXm),n)*mats(dZm,n))==0;
{ 
  % empty code
}
% (4,1) requires trace(mats(Ft*lsqr(Ft,-dXm),n)*mats(dZm,n))==0;
%       requires Ft*lsqr(Ft,-dXm)==-dXm
%       ensures  trace(mats(dXm,n)*mats(dZm,n))==0
{
  % empty code
}
% (5,1) requires trace(mats(dXm,n)*mats(dZm,n))==0;
%       requires trace(Xm*mats(dZm,n))+trace(mats(dXm,n)*Zm)+trace(Xm*Zm)-0.75*trace(Xm*Zm)==0;
%       ensures  trace((Xm+mats(dXm,n))*(Zm+mats(dZm,n)))-0.75*trace(Xm*Zm)==0;
{
  % empty code
}
\end{lstlisting}
\caption{{\tt while} Loop Body: Part III}
\label{annot:loop3}
\end{figure}

In figure~\ref{annot:loop3}, 
the annotations generated are used to prove the formula
\be
\dps Q_{4,1} \wedge Q_{5,1}  \Rightarrow  \Tr{\l X_{-} + \Delta X_{-} \r \l Z_{-} + \Delta Z_{-} \r}-0.75\Tr{X_{-}Z_{-}} = 0, 
\label{strat7}
\ee
in which $Q_{4,1}$ and $Q_{3,2}$ are from figure~\ref{annot:loop3}. 
Each of the Hoare triple represents an individual step in 
the proof of (\ref{strat7}).

We now look at the first contract from 
figure~\ref{annot:loop3}. 
The pre-condition $P_{1,1}$ is generated using the 
skip rule on $Q_{3,1}$ from figure~\ref{annot:loop2}.
The post-condition is generated by noting that 
\annotf{F*lsqr(F,zeros(m,1))} is equal to $\lc 0 \rc^{m}$, 
which can be verified by checking the correctness of
the default contract on the function \annotfc{lsqr}. 

In the next contract, the post-condition is generated by 
applying the transpose operator to 
the first pre-condition statement, which is precisely 
the post-condition of $\mathcal{H}_{1,1}$, 
and then followed by the application of the formula in (\ref{strat0}) using the second 
pre-condition statement in line 7. 
The second pre-condition statement is obtained from the application of the skip rule on 
$Q_{6,1}$ from figure~\ref{annot:init1}. 

The third contract is a proof step that is correct because
\annotfc{mats} and \annotfc{vecs} are unitary transformations. 
Although omitted from the discussion before, 
the properties of unitary transformation can be part of the 
default contract on both of these functions. 

In the fourth contract, the first pre-condition statement is simply the
post-condition of prior contract. The second pre-condition 
statement is inserted by hand.  
Note that its correctness stems from the correctness of the \annotfc{lsqr}. 
If the default contract on \annotfc{lsqr} is satisfied, then
\be
\dps Ft \lsqr{Ft,-dXm}=-dXm 
\label{strat8}
\ee
is true for all $Ft$ and $dXm$. 
By the application of (\ref{strat0}), we get the post-condition in line 19. 

In the last contract, $Q_{4,1}$ and $Q_{4,1}$ from figure~\ref{annot:loop3} are
the pre-condition statements. The post-condition is generated using a two step
heuristics. 
First the pre-conditions are summed, and then followed by an algebraic transformation
into the form in $Q_{5,1}$. 
The generation of $Q_{5,1}$ required far more ad hoc heuristics than 
the other contracts discussed so far. However note that we can split the last 
contract into even smaller steps such as one for the summing and then followed by the one
using the algebraic transformation. 

\subsection{Proving $\phi - 0.76 \phi_{-} <0$ on the code: Part IV}
The annotations and code described in this subsection are from figure~\ref{annot:loop4} unless
stated otherwise. 
\begin{figure}
\begin{lstlisting}
{
  dpm=lsqr(Ft,-dXm); 
  p=pm+dpm
}
% (1,1) require trace((Xm+mats(dXm,n))*(Zm+mats(dZm,n)))-0.75*trace(Xm*Zm)==0; 
%       ensures trace(X*Z)-0.75*trace(Xm*Zm)==0
{
  X=Xm+mats(dXm,n);
  Z=Zm+mats(dZm,n); 
}
% (2,1) requires trace(Xm*Zm)>0;
%       ensures  0.01*trace(Xm*Zm)>0; 
{
  % empty code
}
% (3,1) requires trace(X*Z)-0.75*trace(Xm*Zm)==0
%       requires 0.01*trace(Xm*Zm)>0; 
%       ensures  trace(X*Z)-0.76*trace(Xm*Zm)<0;
{
  % empty code
}
% (4,1) requires trace(X*Z)-0.75*trace(Xm*Zm)==0
%       requires trace(Xm*Zm)>0 && trace(Xm*Zm)<=0.1;
%       ensures  trace(X*Z)>0 && trace(X*Z)<=0.1;

% (4,2) requires trace(X*Z)-0.76*trace(Xm*Zm)<0
%       ensures  trace(X*Z)-0.76*phim<0; 
{
    phim=trace(Xm*Zm);
}
% (5,1) requires trace(X*Z)>0 && trace(X*Z)<=0.1; 
%       ensures  phi>0 && phi<=0.1;

% (5,2) requires trace(X*Z)-0.76*phim<0; 
%       ensures  phi-0.76*phim<0; 

% (5,3) ensures  phi==trace(X*Z);
{
    phi=trace(X*Z); 
    mu=trace(X*Z); 
}
\end{lstlisting}
\caption{{\tt while} Loop Body: Part IV}
\label{annot:loop4}
\end{figure}

The block of code in line 2 and 3 has no contract. We skip ahead to the block of 
code in lines 8 and 9. For $C_{1}$, the pre-condition is a duplication of $Q_{5,1}$ from
figure~\ref{annot:loop3} and the post-condition $Q_{1,1}$ is generated by using the 
backward substitution rule twice i.e. once for each line of code in the block. 

Next, we jump forward to last block of code in the loop body, which is $C_{5}$. 
For $C_{5}$, using the while loop axiom, we insert $Q_{1,1}$ and $Q_{1,2}$
from figure~\ref{annot:loop0} as post-conditions. 
Applying the backward substitution rule on $Q_{5,1}$ and $Q_{5,2}$ 
results in $P_{5,1}$ and $P_{5,2}$ respectively. 
Additionally, we also have $Q_{5,3}$ in order to ensure consistency with
the clause \annotf{phi==trace(X*Z)} in $P_{1,1}$ from figure~\ref{annot:loop1}. 

We now jump backward to $C_{4}$, and we have the post-conditions $Q_{4,1}$ and
$Q_{4,2}$ that are precisely $P_{5,1}$ and $P_{5,2}$. 
The post-condition $Q_{4,1}$ is not affected by the execution of $C_{4}$, so we cannot
apply the Hoare logic rules. 
Instead we propagated the post-conditions $Q_{1,1}$ and $Q_{2,1}$ from
figure~\ref{annot:loop1} forward and then combined them conjunctively to form $P_{4,1}$. 
With the inserted $P_{4,1}$, verifying $\mathcal{H}_{4,1}$ is the same as checking if
\be
\dps P_{4,1} \Rightarrow Q_{4,1} 
\label{strat9}
\ee
holds. 
The formula (\ref{strat9}) can be discharged by 
a theorem prover for example 
in three steps: first we apply the theory
\be
\dps  y-cx =0 \Rightarrow y=cx
\label{strat10}
\ee
on $Q_{1,1}$ and then followed by the application of
\be
\dps  c>0\wedge c<1 \wedge x<=0.1 \wedge x>0 \Rightarrow cx <=0.1 \wedge cx>0, 
\label{strat11}
\ee
on $Q_{2,1}$ from figure~\ref{annot:loop1}. 
Lastly, by applying the formula in (\ref{strat0}) on the conclusions of
(\ref{strat10}) and (\ref{strat11}), we get $y<=0.1 \wedge y>0$ which leads to $Q_{4,1}$. 

The backward substitution rule is applied on $Q_{4,2}$ to get $P_{4,2}$.
Looking back at $C_{1}$, we can see the post-condition $Q_{1,1}$
is not equivalent to $P_{4,2}$.
Since there are no executable code between $Q_{1,1}$ and $P_{4,2}$,
then we have a contradiction unless 
$Q_{1,1} \wedge \bigwedge_{i}  Q_{i} \ \Rightarrow P_{4,2}$, 
in which $Q_{i}$ belongs to the set of all post-conditions of $C_{1}$. 
We resolve this contradiction by inserted the Hoare triples $\mathcal{H}_{i,1},i=2,3$ as a proof of
\be
\dps Q_{1,1} \wedge \Tr{X_{-}Z_{-}}>0\Rightarrow P_{4,2}. 
\label{form1}
\ee
The generation of $\mathcal{H}_{i,1},i=2,3$, 
used ad hoc heuristics just like the annotations in figure~\ref{annot:loop3}. 
As done for (\ref{strat7}), we first decided to split the proof of the formula 
in (\ref{form1}) into two steps.  
In the first step i.e. $\mathcal{H}_{2,1}$, 
we have the pre-condition being \annotf{trace(Xm*Zm)>0}, 
which is true by conjunctive simplification of 
$Q_{2,1}$ from figure~\ref{annot:loop1}, 
and a post-condition $Q_{2,1}$ constructed using the theory
\be
\dps c>0 \wedge x>0 \Rightarrow cx>0,  
\label{strat12}
\ee
with a value of $0.01$ being chosen purposefully for the positive constant $c$. 
In $\mathcal{H}_{3,1}$, we insert $Q_{1,1} \wedge Q_{2,1}$ as the pre-condition and
$P_{4,2}$ as the post-condition. 
The correctness of $\mathcal{H}_{3,1}$ can be verified by applying the theory
\be
\dps y=0 \wedge cx>0 \Rightarrow y-cx<0.
\label{strat13}
\ee

By verifying $P_{4,2}$ , we have now completed the discussion on 
the annotations to assist the mechanical proof of the correctness of the invariants 
\annotf{phi>0 \&\& phi<=0.1} and \annotf{phi-0.76*phim<0}. 
In the next subsection, we discuss, albeit in a far less mechanical way, 
on the possible annotations to prove the correctness of the invariant \annotf{X>0 \&\& Z>0}. 

\subsection{The loop invariant $X\succ 0 \wedge Z\succ 0$}

In the following subsections, 
we describe some of the key annotations used to 
show $Z\succ 0 \wedge X\succ 0$ hold as the invariant of the {\tt while} loop. 
For the sake of brevity, some of the intermediate proof annotations are 
omitted. 

First we claim that 
$X$ and $Z$ are assigned initial values such that
\be
\dps \|XZ-\mu I \|_{F} \leq 0.3105 \mu
\label{central_path0}
\ee
holds. 
\begin{remark}
The property in (\ref{central_path0}) is a constraint on the initial 
set of points that can be assigned to $X$. 
This constraint 
guarantees that $X$ is initialized to within a neighborhood of the
central path, which in turn ensures a good execution time. 
For some optimization problems, deviations from the central path can result in 
bounded but unacceptably large execution time in the real-time context. 
Efficient methods exist in the interior point method literature (see~\cite{lmcgovern}) to 
guarantee initialization within a certain small neighborhood of the central path. 
\end{remark}
We now insert (\ref{central_path0}) and $X \succ 0 \wedge Z\succ 0$ as 
the invariants of the {\tt while} loop. 
\begin{figure}
\begin{lstlisting}
% (1,1) requires norm(X*Z-mu*eye(n,n))<=0.3105*mu && X>0 && Z>0; 
%       ensures  norm(X*Z-mu*eye(n,n))<=0.3105*mu && X>0 && Z>0; 
{
  while (phi>epsilon) do
  	.
	.
	.
  end
}
\end{lstlisting}
\caption{Positive-Definiteness of $X$ and $Z$ as Invariants} 
\label{annot:loop5}
\end{figure}

\subsection{Proving $X\succ 0 \wedge Z\succ0$ on the code: Part I}

All annotations discussed in this subsection are from figure~\ref{annot:loop6} unless
explicitly stated otherwise. 
\begin{figure}
\begin{lstlisting}
% (1,1) requires norm(X*Z-mu*eye(n,n))<=0.3105*mu;
%       requires mu==trace(X*Z)/n;
%       ensures  norm(X*Z-trace(X*Z)/n*eye(n,n))<=0.3105*trace(X*Z)/n;
{
  % empty
}
% (2,1) requires X>0 && Z>0;
%       ensures  Xm>0 && Zm>0;

% (2,2) requires norm(X*Z-trace(X*Z)/n*eye(n,n))<=0.3105*trace(X*Z)/n;
%       ensures  norm(Xm*Zm-trace(Xm*Zm)/n*eye(n,n))<=0.3105*trace(Xm*Zm)/n;
{
  Xm=X
  Zm=Z;
  pm=p;
}
% (3,1) requires norm(Xm*Zm-trace(Xm*Zm)/n*eye(n,n))<=0.3105*trace(Xm*Zm)/n;
%       ensures  norm(Xm*Zm-mu*eye(n,n))<=0.3105*mu; 
{ 
  mu=trace(Xm*Zm)/n;
}
\end{lstlisting}
\caption{Invariant $X\succ0 \wedge Z\succ0$ Part I}
\label{annot:loop6}
\end{figure}
The first pre-condition statement in $\mathcal{H}_{1,1}$ 
is an invariant from $Q_{1,1}$ of figure~\ref{annot:loop5}. 
The second pre-condition statement is obtained from $Q_{10,2}$ from figure~\ref{annot:init2}. 
We apply the formula in~\ref{strat0} to get the post-condition
\be
\dps \|XZ-\frac{\Tr{XZ}}{n} I \|_{F} \leq 0.3105 \frac{\Tr{XZ}}{n}. 
\label{q11}
\ee
Next, for $C_{2}$, we insert two contracts. 
The first contract has the pre-condition $X\succ 0 \wedge Z\succ0$, which
is obtained from $Q_{1,1}$ of figure~\ref{annot:loop5}. 
The post-condition $Q_{2,1}$ is generated by the repeated applications of the 
backward substitution rule.
For the second contract, the pre-condition is simply $Q_{1,1}$ and 
the post-condition is also generated by the repeated application of the
backward substitution rule. 
We now move to $C_{3}$ which has only one contract. The pre-condition
in $\mathcal{H}_{3,1}$ is simply $Q_{2,2}$ and the post-condition $Q_{3,1}$ 
is generated by another application of the backward substitution rule.

\subsection{Proving $X\succ 0 \wedge Z\succ0$ on the code: Part II}
The annotations described in this subsection are from figure~\ref{annot:loop7}
unless explicitly stated otherwise. 
The post-condition $Q_{2,2}$ of figure~\ref{annot:loop5} implies two properties
that are vital to proving the correctness of $Z\succ 0$ and $X\succ 0$. 
The insertion of those properties result in the next two Hoare triples. 
We have $\mathcal{H}_{1,1}$, which is obtained from the formula
\be
\dps P_{1,1} \Rightarrow \|Z_{-}^{-0.5} \Delta Z Z_{-}^{-0.5} \|_{F} \leq 0.7,
\label{pre111}
\ee
with $\Delta Z$ that satisfies the equation in (\ref{taut:0}) 
and $P_{1,1}$ being $Q_{2,2}$ from figure~\ref{annot:loop5}. 
The value $0.7$ in (\ref{pre111}) is obtained from an 
over-approximation of the expression $\frac{ \sqrt{n \l 1-\sigma \r^{2} + 0.3105^2}}{1-0.3105}$. 
The proof for this result is skipped here and can be found in~\cite{Monteiro97}. 
\begin{figure}
\begin{lstlisting}
% (1,1) requires norm(Xm*Zm-mu*eye(n,n))<=0.3105*mu;
%       ensures  norm((Zm^(0.5))^(-1)*mats(lsqr(F,zeros(m,1)),n)*Zm^(0.5))<=0.7;

% (1,2) requires norm(Xm*Zm-mu*eye(n,n))<=0.3105*mu;
%       requires norm(Zm^(0.5))^(-1)*mats(lsqr(krons((Zm^(0.5))^(-1)*Zm,(Zm^(0.5))',n,m),vecs(sigma*mu*eye(n,n)-(Zm^(0.5))*Xm*(Zm^(0.5)))-krons((Zm^(0.5))^(-1),(Zm^(0.5))'*Xm,n,m)*lsqr(F,zeros(m,1))),n)*mats(lsqr(F,zeros(m,1)),n)*Zm^(0.5))<=0.3105*sigma*mu;
{
  % empty
}
% (2,1) requires norm((Zm^(0.5))^(-1)*mats(lsqr(F,zeros(m,1)),n)*Zm^(0.5))<=0.7;
%       ensures  norm(Zhi*mats(dZm,n)*Zhi)<=0.7;

% (2,2) requires norm(Zm^(0.5))^(-1)*mats(lsqr(krons((Zm^(0.5))^(-1)*Zm,(Zm^(0.5))',n,m),vecs(sigma*mu*eye(n,n)-(Zm^(0.5))*Xm*(Zm^(0.5)))-krons((Zm^(0.5))^(-1),(Zm^(0.5))'*Xm,n,m)*lsqr(F,zeros(m,1))),n)*mats(lsqr(F,zeros(m,1)),n)*Zm^(0.5))<=0.3105*sigma*mu;
%       ensures  norm(Zhi*mats(dXm,n)*mats(dXm,n)*Zhi)<=0.3105*sigma*mu;
{
  Zh=Zm^(0.5); 
  Zhi=Zh^(-1); 
  G=krons(Zhi,Zh'*Xm,n,m);
  H=krons(Zhi*Zm,Zh',n,m);
  r=sigma*mu*eye(n,n)-Zh*Xm*Zh;
  dZm=lsqr(F,zeros(m,1));
  dXm=lsqr(H, vecs(r)-G*dZm);
}
% (3,1) requires mats(H*dXm)+mats(G*dZm)==sigma*mu*eye(n,n)-Zh*Xm*Zh; 
%       requires norm(Zhi*mats(dXm,n)*mats(dXm,n)*Zhi)<=0.3105*sigma*mu;
%       ensures  0.5*norm(Zhi*((Zm+mats(dZm,n)*(Xm+mats(dXm,n)-sigma*mu*eye(n,n))*Zh+Zh'*((Xm+mats(dXm,n)*(Zm+mats(dZm,n)-sigma*mu*eye(n,n))*Zhi)<=0.3105*sigma*mu;
{
  % empty
}
  dpm=lsqr(Ft,-dXm);
  p=pm+dpm;
% (4,1) requires norm(Zhi*mats(dZm,n)*Zhi)<=0.7;
%       ensures  Zm+mats(dZm,n)>0;
{ 
  % empty
}
\end{lstlisting}
\caption{Annotations for the Invariant $X\succ0 \wedge Z\succ0$: Part II}
\label{annot:loop7}
\end{figure}
We also have $\mathcal{H}_{1,2}$, which is obtained from the formula
\be
\dps P_{2,2} \Rightarrow \|Z_{-}^{-0.5} \Delta X\Delta Z Z_{-}^{0.5}\|_{F} \leq 0.3105 \sigma \mu,
\label{pre122}
\ee
with $\Delta X$ and $\Delta Z$ that satisfy (\ref{taut:0}) and (\ref{taut:1})
and $P_{1,2}$ also being $Q_{2,2}$ from figure~\ref{annot:loop5}. 

Next, the block of code $C_{2}$ computes the Newton directions $\Delta X_{-}$ and $\Delta Z_{-}$. 
We duplicate $Q_{1,1}$ and $Q_{1,2}$ to form $P_{2,1}$ and $P_{2,2}$ respectively. 
Using the backward substitution rule on $Q_{2,1}$, we get the post-condition
\be
\dps \|Z_{-}^{-0.5} \Delta Z_{-} Z_{-}^{-0.5} \|_{F} \leq 0.7.
\label{post01}
\ee
Applying the backward substitution rule again on $P_{2,2}$, we get the post-condition
\be
\dps \|Z_{-}^{-0.5} \Delta X_{-} \Delta Z_{-} Z_{-}^{0.5}\|_{F} \leq 0.3105 \sigma \mu, 
\label{post02}
\ee

We move to the next Hoare triple which is $\mathcal{H}_{3,1}$. 
The pre-condition statement 
\annotf{mats(H*dXm)+mats(G*dZm)==sigma*mu*eye(n,n)-Zh*Xm*Zh} is correct by noting 
that \annotf{mats(H*dXm)} is equal to \annotf{mats(vecs(r)-G*dXm)} because of the 
assignment in line 21, and \annotf{mats(vecs(r)-G*dXm)} reduces to \annotf{r-mats(G*dXm)}. 
An equivalent formula to \annotf{mats(H*dXm)+mats(G*dZm)==sigma*mu*eye(n,n)-Zh*Xm*Zh} is
\be
\begin{split}
	\dps 0.5 \l Z_{-}^{-0.5} \l \Delta Z_{-} X_{-} + Z_{-} \Delta X_{-} \r Z_{-}^{0.5} + Z_{-}^{0.5} \l X_{-} \Delta Z_{-} + \Delta X_{-} Z_{-} \r Z_{-}^{-0.5} \r\\ 
         =  \sigma \mu I - Z_{-}^{0.5} X_{-} Z_{-}^{0.5}.
\end{split}
\label{post03}
\ee
The second pre-condition statement of $\mathcal{H}_{3,1}$ is $Q_{2,2}$ unchanged. 
Given the pre-condition statements, we can deductively arrive at the post-condition $Q_{3,1}$, which 
is
\be
\begin{split}
\dps 0.5 \| Z_{-}^{-0.5} \l \l Z_{-} + \Delta Z_{-} \r \l X_{-} + \Delta X_{-} \r - \sigma \mu I \r Z_{-}^{0.5} + \\ 
Z_{-}^{0.5} \l \l X_{-} + \Delta X_{-} \r \l Z_{-} + \Delta Z_{-} \r - \sigma \mu I \r Z_{-}^{-0.5} \|_{F}  \\ 
 \leq 0.3105 \sigma \mu. 
 \end{split}
\label{pre00}
\ee
\begin{remark}
Note that in (\ref{pre00}), we use the 
expression $Z_{-}^{0.5}$ to represent the variable \annotc{Zh}
and the expression $Z_{-}^{-0.5}$ to represent \annotc{Zhi}. 
This can be done because of the assignments in lines 15 and 16. 
For the sake of brevity, we do not annotate the steps that lead to (\ref{pre00}) as Hoare triples on the 
code, nonetheless here we give a sketch of the constructive proof. 
First note that
\be
\begin{split}
\dps 0.5 \l Z_{-}^{-0.5} \l \Delta Z_{-} X_{-} + Z_{-} \Delta X_{-} \r Z_{-}^{0.5}+ Z_{-}^{0.5} \l X_{-} \Delta Z_{-} + \Delta X_{-} Z_{-} \r Z_{-}^{-0.5} \right.+ \\
\left. Z_{-}^{0.5} X_{-} Z_{-}^{0.5} - \sigma \mu I  +  Z_{-}^{-0.5} \l \Delta Z_{-} \Delta X_{-} \r Z_{-}^{0.5} +  Z_{-}^{0.5} \l \Delta X_{-} \Delta Z_{-}  \r Z_{-}^{-0.5} \r\\
=\dps 0.5 \l Z_{-}^{-0.5} \l \l Z_{-} + \Delta Z_{-} \r \l X_{-} + \Delta X_{-} \r - \sigma \mu I \r Z_{-}^{0.5} + \right.\\
\left.Z_{-}^{0.5} \l \l X_{-} + \Delta X_{-} \r \l Z_{-} + \Delta Z_{-} \r - \sigma \mu  I \r Z_{-}^{-0.5}\r . 
\end{split}
\label{pre01}
\ee
holds because the left hand side is an algebraic expansion of the right hand side. 
Second, apply the formula in (\ref{strat0}) on the conjunction of (\ref{pre01}) and (\ref{post03}), 
(\ref{pre01}) is reduced to
\be
\begin{split}
\dps  Z_{-}^{-0.5} \l \Delta Z_{-} \Delta X_{-} \r Z_{-}^{0.5} +  Z_{-}^{0.5} \l \Delta X_{-} \Delta Z_{-}  \r Z_{-}^{-0.5} \\
=\dps 0.5 \l Z_{-}^{-0.5} \l \l Z_{-} + \Delta Z_{-} \r \l X_{-} + \Delta X_{-} \r - \sigma \mu I \r Z_{-}^{0.5} + \right.\\
\left.Z_{-}^{0.5} \l \l X_{-} + \Delta X_{-} \r \l Z_{-} + \Delta Z_{-} \r - \sigma \mu  I \r Z_{-}^{-0.5}\r . 
\end{split}
\label{pre02}
\ee
Finally, 
apply the transitive property of the comparison operators
to $Q_{2,2}$ and the Frobenius norm of (\ref{pre02}), we get
the post-condition in (\ref{pre00}). 
\end{remark} 

Now we move ahead to the last Hoare triple of figure~\ref{annot:loop7}, 
which is $\mathcal{H}_{4,1}$. 
This Hoare triple is generated using the formula
\be
\dps Q_{2,1} \Rightarrow Z_{-} + \Delta Z_{-} \succ 0,
\label{post221}
\ee
\begin{remark}
The formula in (\ref{post221}) is correct and to see that, 
note 
\be
\dps Q_{2,1} \Rightarrow \|Z_{-}^{-0.5} \Delta Z_{-} Z_{-}^{-0.5} \|_{F} < 1 \Rightarrow I + Z_{-}^{-0.5} \Delta Z_{-} Z_{-}^{-0.5} \succ 0.
\label{cond01}
\ee
and that 
\be
\dps I + Z_{-}^{-0.5} \Delta Z_{-} Z_{-}^{-0.5} = Z_{-}^{-0.5} \l Z_{-} + \Delta Z_{-} \r Z_{-}^{-0.5}. 
\label{cond02}
\ee
\end{remark} 

\subsection{Proving $X\succ 0 \wedge Z\succ0$ on the code: Part III}
Now we move forward to the last part of the loop body. 
The annotations described in this subsection are from figure~\ref{annot:loop8}
unless explicitly stated otherwise. 
\begin{figure}[htp]
\begin{lstlisting}
% (1,1) requires 0.5*norm(Zhi*((Zm+mats(dZm,n)*(Xm+mats(dXm,n)-sigma*mu*eye(n,n))*Zh+Zh'*((Xm+mats(dXm,n)*(Zm+mats(dZm,n)-sigma*mu*eye(n,n))*Zhi')<=0.3105*sigma*mu;
%       ensures  0.5*norm(Zhi*(X*Z-sigma*mu*eye(n,n))*Zh+Zh'*(Z*X-sigma*mu*eye(n,n))*Zhi')<=0.3105*sigma*mu; 

% (1,2) requires Zm+mats(dZm,n)>0;
%       ensures  Z>0; 
{
  X=Xm+mats(dXm,n);
  Z=Zm+mats(dZm,n);
}
% (2,1) requires Z>0 && Zm>0; 
%       ensures  norm(Z^(0.5)*X*Z^(0.5)-sigma*mu*eye(n,n))<=0.5*norm(Zhi*(X*Z-sigma*mu*I)*Zh+Zh'*(Z*X-sigma*mu*I)*Zhi');  
{
  % empty
}
  phim=trace(Xm*Zm);
  phi=trace(X*Z);
% (3,1) requires trace(X*Z)-sigma*trace(Xm*Zm)==0;
%       requires trace(Xm*Zm)==n*mu;
%       ensures  trace(X*Z)-sigma*n*mu==0;
{
  % empty
}
% (4,1) requires trace(X*Z)-sigma*n*mu==0; 
%       requires norm(Z^(0.5)*X*Z^(0.5)-sigma*mu*eye(n,n))<=0.5*norm(Zhi*(X*Z-sigma*mu*I)*Zh+Zh'*(Z*X-sigma*mu*eye(n,n))*Zhi');
%       requires 0.5*norm(Zhi*(X*Z-sigma*mu*eye(n,n))*Zh+Zh'*(Z*X-sigma*mu*eye(n,n))*Zhi')<=0.3105*sigma*mu;
%       ensures  norm(Z^(0.5)*X*Z^(0.5)-trace(X*Z)/n*eye(n,n))<=0.3105*trace(X*Z)/n;
{
  % empty
}
% (5,1) requires norm(Z^(0.5)*X*Z^(0.5)-trace(X*Z)/n*eye(n,n))<=0.3105*trace(X*Z)/n;
%       ensures  norm(Z^(0.5)*X*Z^(0.5)-mu*eye(n,n))<=0.3105*mu;
{
  mu=trace(X*Z)/n;
}
% (6,1) requires norm(Z^(0.5)*X*Z^(0.5)-mu*eye(n,n))<=0.3105*mu;
%       ensures  norm(XZ-mu*eye(n,n))<=0.3105*mu;

% (6,2) requires Z>0;
%       requires norm(Z^(0.5)*X*Z^(0.5)-mu*eye(n,n))<=0.3105*mu;
%       ensures  X>0; 
{
  % empty
}
\end{lstlisting}
\caption{Annotations for the Invariant $X\succ0 \wedge Z\succ0$: Part III}
\label{annot:loop8}
\end{figure}
We insert two contracts for the block of code $C_{1}$. 
The pre-conditions of the two contracts are respectively $Q_{3,1}$ and $Q_{4,1}$ 
of figure~\ref{annot:loop7} propagated forward using the skip rule. 
By the application of the backward substitution, we obtain the post-conditions 
$P_{1,1}$ and $P_{1,2}$. 
The post-condition $Z\succ 0$ completes the proof for a part of the invariant
$Q_{1,1}$ from figure~\ref{annot:loop5}. 
We still have to show  $X\succ 0$ and $\|XZ-\mu I \|_{F} \leq 0.3105 \mu$ hold. 

The next Hoare triple, which is $\mathcal{H}_{2,1}$, is generated using the formula,  
$\forall Z\succ 0, Z_{-}\succ 0$, 
\be
\begin{split} 
\dps \| Z^{0.5} X Z^{0.5} - \sigma \mu I \|_{F}  \leq \frac{1}{2} \| Z_{-}^{-0.5} \l Z X - \sigma \mu I \r Z_{-}^{0.5}  + \\ 
Z_{-}^{0.5} \l X Z - \sigma \mu I\r Z_{-}^{-0.5} \|_{F}
\end{split}
\label{pre04}
\ee 
holds. 
The post-condition of $\mathcal{H}_{2,1}$ is precisely the statement in (\ref{pre04})
with \annotc{Zh} being $Z_{-}^{0.5}$ and \annotc{Zhi} being $Z_{-}^{-0.5}$. 

Now we forward again to $\mathcal{H}_{3,1}$.  
Recall the post-condition $Q_{1,1}$ from figure~\ref{annot:loop4}, which
is \annotf{trace(X*Z) - 0.75*trace(Xm*Zm)==0}. Also recall that the post-condition $Q_{1,1}$
from figure~\ref{annot:init2}, which is \annotf{sigma==0.75}. 
We can apply the rule in (\ref{strat0}) to \annotf{trace(X*Z) - 0.75*trace(Xm*Zm)==0 \&\& sigma==0.75}, 
and obtain the first pre-condition statement \annotf{trace(X*Z)-sigma*trace(Xm*Zm)==0}. 
By propagating forward $Q_{1,1}$ from figure~\ref{annot:loop1}, we get the second pre-condition statement. 
Apply again the rule from (\ref{strat0}) on $P_{3,1}$, 
we get the post-condition \annotf{trace(X*Z)-sigma*n*mu==0}. 

In the next Hoare triple, which is $\mathcal{H}_{4,1}$, the pre-condition is formed by 
combining the post-conditions $Q_{i,1},i=1,2,3$. 
The post-condition $Q_{4,1}$ is generated by first 
noting that due to the transitivity of \annotc{<=}, the formula
\be
\dps Q_{2,1} \wedge Q_{1,1} \Rightarrow \| Z^{0.5} X Z^{0.5} - \sigma \mu I \|_{F} \leq 0.3105 \mu
\label{post413}
\ee
holds. 
Second, apply the rule in (\ref{strat0}) to the conjunction of (\ref{post413}) and $Q_{3,1}$, you
get precisely the post-condition $Q_{4,1}$. 

Finally we move forward to the block of code in line 33, in which
the variable \annotc{mu} is updated with the expression $\frac{\Tr{XZ}}{n}$. 
Apply the backward substitution rule on the pre-condition, which is simply 
$Q_{4,1}$ unaltered, we get a post-condition of
\be
\dps \| Z^{0.5} X Z^{0.5} - \mu I \|_{F} \leq 0.3105 \mu. 
\label{post05}
\ee

The post-condition in (\ref{post05}) i.e. $Q_{5,1}$ is used to generate
the last two Hoare triples of figure~\ref{annot:loop8}. 
The Hoare triple $\mathcal{H}_{6,1}$ is a result of the formula
\be
\dps  \| Z^{0.5} X Z^{0.5} - \mu I \|_{F} \leq 0.3105 \mu \Rightarrow  \| XZ -\mu I \|_{F} \leq 0.3105 \mu, 
\label{post06}
\ee
which we know is correct beforehand. 
For $\mathcal{H}_{6,2}$, we have the pre-condition $Z\succ 0 \wedge Q_{5,1}$ which implies $X\succ0$, 
and thus completes the annotation process.

\section{Autocoding of Convex Optimization Algorithms and Automatic Verification}

In this section, we describe some general ideas towards the credible autocoding of convex optimization algorithms. 
In credible autocoding, the optimization semantics, such as those described in the manual process 
in the previous section, is generated automatically. 
The variations of the interior point method discussed in this paper 
is relatively simple with changes in one of the parameters such as the symmetrizing
scaling matrix $T$, the step size $\alpha$ which is defaulted to $1$ in the algorithm description, 
the duality gap reduction parameter $\sigma$, etc. 
The more complex interior point implementations such as those with heuristics in the predictor, 
can also be specified using a few additional parameters. 
Since many optimization programs differs from each other only in a finite set of 
parameter, and the input problem, 
we proppose an autocoding approach based on a set of standard parameterized mappings
of interior-point algorithms to the output code. 
The same approach applies to the generation of the optimization semantics.
We can construct a set of mappings from the type of interior point algorithm 
to a set of standard paramaterized optimizaiton semantics i.e. such as the ones
described in the previous section. 
The autocoding process, roughly speaking, becomes a procedure to choose the mapping, 
followed by insertion of the pre-defined 
semantics into the generated code, and then subsituting in the values 
of the parameters and the input problem. 

As we discussed briefly before, for domain-specific properties such as $Z\succ 0 \wedge X\succ 0$, 
the annotations are obtained usually from the steps of a complex
proof that cannot always be discharge by a generic automated proof-checkers. 
The automatic verification of the annotated output most likely would require
a theorem prover based tool. 
The theorem prover need to be adapted to handle the theories 
and formulas described in the previous section.
The current theories available in the theorem prover PVS 
are not equipped to handle a lot of annotations discussed, especially in
figures~\ref{annot:loop7} and~\ref{annot:loop8}. 
The feasible approach would be to use the same method that 
we took for control system invariants i.e. to construct a few key theorems, 
which can be repeatedly applied to many autocoded optimization programs as 
they only differs from each other in the input data, step size or the Newton direction. 
Key theorems including properties on Frobenius norms, matrix operator theory, etc.

\section{Future Work}

In this paper, we introduce an approach to
communicate high-level functional properties of convex optimization algorithms and their proofs down to the code level. 
Now we want to discuss several possible directions of interest that one can explore in the future. 
On the more theoretical front, we can look at the
possibility that there might exist linear approximations to the potential function used in the construction of the invariant. 
Having linear approximations would possibly allow us to construct efficient automatic 
decision procedures to verify the annotations on the code level. 
On the more practical front, we also need to demonstrate the expression of the 
interior point semantics on an implementation-level language like C rather than the high-level computational language
used in this paper. 
Related to that is the construction of a prototype tool that is capable of autocoding a variety of
convex optimization programs along with their proofs down to the code level.  
There is also a need to explore the verification of those proof annotations on the code level.
It is clear that none of the Hoare triple annotations shown in the previous section, even 
expressed in a more realistic annotation language, can be handled by existing verification tools.
Finally, we also need to be able to reason about the invariants introduced in this paper in the presence of the 
numerical errors due to floating-point computations. 

\section{Conclusions} 

This paper proposes the transformation of high-level functional properties of interior point method algorithms down to implementation level for 
certification purpose. 
The approach is taken from a previous work done for control systems. 
We give an example of a primal-dual interior point algorithms and its convergence property. 
We show that the high-level proofs can be used as annotations for the verification of an online optimization program. 

\section{Acknowledgements}

The authors would like to acknowledge support from the V\'erification de l'Optimisation Rapide Appliqu\'ee \`a la Commande Embarqu\'ee (VORACE) project, 
the NSF Grant CNS - 1135955 “CPS: Medium: Collaborative Research: Credible Autocoding and Verification of Embedded Software (CrAVES)” and 
Army Research Office's MURI Award W911NF-11-1-0046

\bibliographystyle{abbrv}
\bibliography{complete} 
\section{Appendix}

\subsection{Vectorization Functions}
The function \annotc{vecs} is similar to the standard vectorization function but 
specialized for symmetric matrices. 
It is defined as, for $1\leq i < j \leq n$ and $M \in \mathbb{S}^{n}$, 
\be
\dps \vecs{M} = \begin{bmatrix} M_{11}, \ldots, &\sqrt{2} M_{ij},& \ldots,& M_{nn}  \end{bmatrix}^{\m{T}}. 
\label{m1}
\ee
The factor $\sqrt{2}$ ensures the function \annotc{vecs} preserves the distance defined by the 
respective inner products of $\mathbb{S}^{n}$ and $\R^{\frac{n \l n+1\r}{2}}$. 
The function \annotc{mats} is the inverse of \annotc{vecs}. 
The function \annotc{krons}, denoted by the symbol $\krons$, 
is similar to the standard Kronecker product but specialized for symmetric 
matrix equations. 
It has the property
\be
\dps \l Q_{1} \krons Q_{2} \r \vecs{(M)} = \vecs{\l \frac{1}{2} \l Q_{1} M Q_{2}^{\m{T}} + Q_{2} M Q_{1}^{\m{T}} \r \r}. 
\label{krons}
\ee
Let $Q_{1}=TZ$ and $Q_{2}=T_{inv}$ and $M=\Delta X$, we get
\be
\dps \l TZ \krons T_{inv} \r \vecs{(\Delta X)} = \vecs{\l \frac{1}{2} \l TZ \Delta X T_{inv} + T_{inv} \Delta X ZT\r \r}. 
\label{krons1}
\ee
Additionally, let $Q_{1}=T$, $Q_{2}=XT_{inv}$, and $M=\Delta Z$, we get
\be
\dps \l T  \krons XT_{inv} \r \vecs{(\Delta Z)} = \vecs{\l \frac{1}{2} \l T \Delta Z XT_{inv} + T_{inv} X \Delta Z T\r \r}. 
\label{krons2}
\ee
Combining (\ref{krons1}) and (\ref{krons2}), we get exactly the left hand side of the third equation in (\ref{newton12}). 
Given a $\Delta Z$, we can compute $\Delta X$ by solving $Ax=b$ for $x$ where
\be\ba{c}
A=\l TZ \krons T_{inv} \r \cr
\Delta X = \mats{\l x \r} \cr
b=\vecs{\l \sigma \mu I - T_{inv} X T_{inv} \r} - \l T  \krons XT_{inv} \r \vecs{(\Delta Z)}. 
\ea
\label{krons3}
\ee

\subsection{Annotated Code}
\label{sec:annot}
\begin{lstlisting}
%% Example SDP Code: Primal-Dual Short-Step Algorithm
% ensures F0>0; 
{
  F0=[1, 0; 0, 0.1];
}
% ensures transpose(F1)==F1; 
{ 
  F1=[-0.750999 0.00499; 0.00499 0.0001];
}
% ensures transpose(F2)==F2;
{
  F2=[0.03992 -0.999101; -0.999101 0.00002];
}
% ensures transpose(F3)==F3;
{
  F3=[0.0016 0.00004; 0.00004 -0.999999];
}
% ensures smat(b)>0; 
{
  b=[0.4; -0.2; 0.2];
}
F=[vecs(F1); vecs(F2); vecs(F3)];
% ensures Ft==transpose(F);
{
  Ft=F'; 
}
% ensures n>=1;
{ 
  n=length(F0);
}
% ensures m>=1;
{
  m=length(b);
}
% ensures Z>0; 
{
  Z=mats(lsqr(F,-b),n); 
}
% ensures X>0 

% ensures trace(X*Z)<=0.1; 
{
  X=[0.3409 0.2407; 0.2407 0.9021];
}
% requires Z>0 && X>0; 
% ensures  trace(X*Z)>0;
{
  % empty code
}
% ensures transpose(P)==P; 
{
  P=mats(lsqr(Ft,vecs(-X-F0)),n);
  p=vecs(P);
}
% ensures epsilon>0
{
  epsilon=1e-8;
}
% ensures sigma==0.75;
{
  sigma=0.75;
}
% requires trace(X*Z)<=0.1;
% ensures  phi<=0.1; 

% requires trace(X*Z)>0; 
% ensures  phi>0; 

% ensures  phi==trace(X*Z); 
{
  phi=trace(X*Z); 
}
% requires phi>0;
% ensures  phi-0.76/0.75*phi<0; 
{
  % empty code
}
% requires phi-0.76/0.75*phi<0; 
% ensures  phi-0.76*phim<0; 
{
  phim=1/0.75*phi; 
}
% ensures  mu==trace(X*Z)/n;
{
  mu=trace(X*Z)/n;
}

% requires phi>0 && phi<=0.1;
% ensures  phi>0 && phi<=0.1;

% requires phi-0.76*phim<0; 
% ensures  phi-0.76*phim<0; 

% requires norm(X*Z-mu*eye(n,n))<=0.3105*mu && X>0 && Z>0; 
% ensures  norm(X*Z-mu*eye(n,n))<=0.3105*mu && X>0 && Z>0; 
{
  while (phi>epsilon) do
  % requires phi>0 && phi<=0.1 && phi==trace(X*Z); 
  % ensures  trace(X*Z)>0 && trace(X*Z)<=0.1; 

  % requires norm(X*Z-mu*eye(n,n))<=0.3105*mu;
  % requires mu==trace(X*Z)/n;
  % ensures  norm(X*Z-trace(X*Z)/n*eye(n,n))<=0.3105*trace(X*Z)/n;
  
  {
    % empty code
  }
  % requires trace(X*Z)>0 && trace(X*Z)<=0.1; 
  % ensures  trace(Xm*Zm)>0 && trace(Xm*Zm)<=0.1; 

  %  requires X>0 && Z>0;
  %  ensures  Xm>0 && Zm>0;
  
  %  requires norm(X*Z-trace(X*Z)/n*eye(n,n))<=0.3105*trace(X*Z)/n;
  %  ensures  norm(Xm*Zm-trace(Xm*Zm)/n*eye(n,n))<=0.3105*trace(Xm*Zm)/n;
  {
    Xm=X;
    Zm=Z;
  }
    pm=p;
  % requires n>=1; 
  % ensures  trace(Xm*Zm)==n*mu; 

  %  requires norm(Xm*Zm-trace(Xm*Zm)/n*eye(n,n))<=0.3105*trace(Xm*Zm)/n;
  %  ensures  norm(Xm*Zm-mu*eye(n,n))<=0.3105*mu;
  {
    mu=trace(Xm*Zm)/n;
  }
  %  requires norm((Zm^(0.5))^(-1)*mats(lsqr(F,zeros(m,1)),n)*Zm^(0.5))<=0.7;
  %  ensures  norm(Zhi*mats(dZm,n)*Zhi)<=0.7;
  
  %  requires norm(Zm^(0.5))^(-1)*mats(lsqr(krons((Zm^(0.5))^(-1)*Zm,(Zm^(0.5))',n,m),vecs(sigma*mu*eye(n,n)-(Zm^(0.5))*Xm*(Zm^(0.5)))-krons((Zm^(0.5))^(-1),(Zm^(0.5))'*Xm,n,m)*lsqr(F,zeros(m,1))),n)*mats(lsqr(F,zeros(m,1)),n)*Zm^(0.5))<=0.3105*sigma*mu;
  %  ensures  norm(Zhi*mats(dXm,n)*mats(dXm,n)*Zhi)<=0.3105*sigma*mu;
  {
    % requires n*mu==trace(Xm*Zm); 
    % requires trace(Xm*mats(lsqr(F,zeros(m,1)),n))+trace(mats(lsqr(krons((Zm^(0.5))^(-1)*Zm,(Zm^(0.5))',n,m),vecs(sigma*mu*eye(n,n)-(Zm^(0.5))*Xm*(Zm^(0.5)))-krons((Zm^(0.5))^(-1),(Zm^(0.5))'*Xm,n,m)*lsqr(F,zeros(m,1))),n)*Zm)+trace(Xm*Zm)-sigma*n*mu==0; 
    % ensures  trace(Xm*mats(lsqr(F,zeros(m,1)),n))+trace(mats(lsqr(krons((Zm^(0.5))^(-1)*Zm,(Zm^(0.5))',n,m),vecs(sigma*mu*eye(n,n)-(Zm^(0.5))*Xm*(Zm^(0.5)))-krons((Zm^(0.5))^(-1),(Zm^(0.5))'*Xm,n,m)*lsqr(F,zeros(m,1))),n)*Zm)+trace(Xm*Zm)-sigma*trace(Xm*Zm)==0;
    {
      % empty
    }
    % requires trace(Xm*mats(lsqr(F,zeros(m,1)),n))+trace(mats(lsqr(krons((Zm^(0.5))^(-1)*Zm,(Zm^(0.5))',n,m),vecs(sigma*mu*eye(n,n)-(Zm^(0.5))*Xm*(Zm^(0.5)))-krons((Zm^(0.5))^(-1),(Zm^(0.5))'*Xm,n,m)*lsqr(F,zeros(m,1))),n)*Zm)+trace(Xm*Zm)-sigma*trace(Xm*Zm)==0; 
    % ensures  trace(Xm*mats(lsqr(F,zeros(m,1)),n))+trace(mats(lsqr(H,vecs(sigma*mu*eye(n,n)-Zh*Xm*Zh)-G*lsqr(F,zeros(m,1))),n)*Zm)+trace(Xm*Zm)-sigma*trace(Xm*Zm)==0; 
    {
      Zh=Zm^(0.5); 
      Zhi=Zh^(-1); 
      G=krons(Zhi,Zh'*Xm,n,m);
      H=krons(Zhi*Zm,Zh',n,m);
    }
    % requires trace(Xm*mats(lsqr(F,zeros(m,1)),n))+trace(mats(lsqr(H,vecs(sigma*mu*eye(n,n)-Zh*Xm*Zh)-G*lsqr(F,zeros(m,1))),n)*Zm)+trace(Xm*Zm)-sigma*trace(Xm*Zm)==0;
    % ensures  trace(Xm*mats(dZm,n))+trace(mats(dXm,n)*Zm)+trace(Xm*Zm)-sigma*trace(Xm*Zm)==0;
    
    % ensures  dZm==lsqr(F,zeros(m,1));
    {
      r=sigma*mu*eye(n,n)-Zh*Xm*Zh;
      dZm=lsqr(F,zeros(m,1));
      dXm=lsqr(H, vecs(r)-G*dZm);
    }
  }
  % requires sigma==0.75; 
  % requires trace(Xm*mats(dZm,n))+trace(mats(dXm,n)*Zm)+trace(Xm*Zm)-sigma*trace(Xm*Zm)==0;
  % ensures  trace(Xm*mats(dZm,n))+trace(mats(dXm,n)*Zm)+trace(Xm*Zm)-0.75*trace(Xm*Zm)==0;
  { 
    % empty code
  }
  % requires  dZm==lsqr(F,zeros(m,1));
  % ensures   lsqr(Ft,-dXm)'*F*dZm==0;
  { 
    % empty code
  }
  % requires lsqr(Ft,-dXm)*F*dZm==0;
  % requires transpose(F)==Ft; 
  % ensures  dot(Ft*lsqr(Ft,-dXm),dZm)==0;
  {
    % empty code
  }
  % requires dot(Ft*lsqr(Ft,-dXm),dZm)==0;  
  % ensures  trace(mats(Ft*lsqr(Ft,-dXm),n)*mats(dZm,n))==0;
  { 
    % empty code
  }
  % requires trace(mats(Ft*lsqr(Ft,-dXm),n)*mats(dZm,n))==0;
  % requires Ft*lsqr(Ft,-dXm)==-dXm
  % ensures  trace(mats(dXm,n)*mats(dZm,n))==0
  {
    % empty code
  }
  % requires trace(mats(dXm,n)*mats(dZm,n))==0;
  % requires trace(Xm*mats(dZm,n))+trace(mats(dXm,n)*Zm)+trace(Xm*Zm)-0.75*trace(Xm*Zm)==0;
  % ensures  trace((Xm+mats(dXm,n))*(Zm+mats(dZm,n)))-0.75*trace(Xm*Zm)==0;
  {
    % empty code
  }
  %  requires mats(H*dXm)+mats(G*dZm)==sigma*mu*eye(n,n)-Zh*Xm*Zh; 
  %  requires norm(Zhi*mats(dXm,n)*mats(dXm,n)*Zhi)<=0.3105*sigma*mu;
  %  ensures  0.5*norm(Zhi*((Zm+mats(dZm,n)*(Xm+mats(dXm,n)-sigma*mu*eye(n,n))*Zh+Zh'*((Xm+mats(dXm,n)*(Zm+mats(dZm,n)-sigma*mu*eye(n,n))*Zhi)<=0.3105*sigma*mu;
  {
    % empty
  }
  {
    dpm=lsqr(Ft,-dXm); 
    p=pm+dpm
  }
  %  requires norm(Zhi*mats(dZm,n)*Zhi)<=0.7;
  %  ensures  Zm+mats(dZm,n)>0;
  { 
    % empty
  }
  % require trace((Xm+mats(dXm,n))*(Zm+mats(dZm,n)))-0.75*trace(Xm*Zm)==0; 
  % ensures trace(X*Z)-0.75*trace(Xm*Zm)==0;

  %  requires 0.5*norm(Zhi*((Zm+mats(dZm,n)*(Xm+mats(dXm,n)-sigma*mu*eye(n,n))*Zh+Zh'*((Xm+mats(dXm,n)*(Zm+mats(dZm,n)-sigma*mu*eye(n,n))*Zhi')<=0.3105*sigma*mu;
  %  ensures  0.5*norm(Zhi*(X*Z-sigma*mu*eye(n,n))*Zh+Zh'*(Z*X-sigma*mu*eye(n,n))*Zhi')<=0.3105*sigma*mu; 
  
  %  requires Zm+mats(dZm,n)>0;
  %  ensures  Z>0; 
  {
    X=Xm+mats(dXm,n);
    Z=Zm+mats(dZm,n); 
  }
  % requires trace(Xm*Zm)>0;
  % ensures  0.01*trace(Xm*Zm)>0; 
  {
    % empty code
  }
  % requires trace(X*Z)-0.75*trace(Xm*Zm)==0
  % requires 0.01*trace(Xm*Zm)>0; 
  % ensures  trace(X*Z)-0.76*trace(Xm*Zm)<0;
  {
    % empty code
  }
  %  requires Z>0 && Zm>0; 
  %  ensures  norm(Z^(0.5)*X*Z^(0.5)-sigma*mu*eye(n,n))<=0.5*norm(Zhi*(X*Z-sigma*mu*I)*Zh+Zh'*(Z*X-sigma*mu*I)*Zhi');  
  {
    % empty
  }
  % requires trace(X*Z)-0.75*trace(Xm*Zm)==0
  % requires trace(Xm*Zm)>0 && trace(Xm*Zm)<=0.1;
  % ensures  trace(X*Z)>0 && trace(X*Z)<=0.1;
  
  % requires trace(X*Z)-0.76*trace(Xm*Zm)<0
  % ensures  trace(X*Z)-0.76*phim<0; 
  {
      phim=trace(Xm*Zm);
  }
  % requires trace(X*Z)>0 && trace(X*Z)<=0.1; 
  % ensures  phi>0 && phi<=0.1;
  
  % requires trace(X*Z)-0.76*phim<0; 
  % ensures  phi-0.76*phim<0; 
  
  % ensures  phi==trace(X*Z);
  {
      phi=trace(X*Z); 
  }
  %  requires trace(X*Z)-sigma*trace(Xm*Zm)==0;
  %  requires trace(Xm*Zm)==n*mu;
  %  ensures  trace(X*Z)-sigma*n*mu==0;
  {
    % empty
  }
  %  requires trace(X*Z)-sigma*n*mu==0; 
  %  requires norm(Z^(0.5)*X*Z^(0.5)-sigma*mu*eye(n,n))<=0.5*norm(Zhi*(X*Z-sigma*mu*I)*Zh+Zh'*(Z*X-sigma*mu*eye(n,n))*Zhi');
  %  requires 0.5*norm(Zhi*(X*Z-sigma*mu*eye(n,n))*Zh+Zh'*(Z*X-sigma*mu*eye(n,n))*Zhi')<=0.3105*sigma*mu;
  %  ensures  norm(Z^(0.5)*X*Z^(0.5)-trace(X*Z)/n*eye(n,n))<=0.3105*trace(X*Z)/n;
  {
    % empty
  }
  %  requires norm(Z^(0.5)*X*Z^(0.5)-trace(X*Z)/n*eye(n,n))<=0.3105*trace(X*Z)/n;
  %  ensures  norm(Z^(0.5)*X*Z^(0.5)-mu*eye(n,n))<=0.3105*mu;
  {
    mu=trace(X*Z)/n;
  }
  %  requires norm(Z^(0.5)*X*Z^(0.5)-mu*eye(n,n))<=0.3105*mu;
  %  ensures  norm(XZ-mu*eye(n,n))<=0.3105*mu;
  
  %  requires Z>0;
  %  requires norm(Z^(0.5)*X*Z^(0.5)-mu*eye(n,n))<=0.3105*mu;
  %  ensures  X>0; 
  {
    % empty
  }
   end
}
\end{lstlisting}

\end{document}